\documentclass[10.5pt]{article}
\usepackage[a4paper,margin=2.3cm]{geometry}
\usepackage{amsmath,amssymb}
\usepackage{graphicx}
\usepackage{booktabs}
\usepackage{makecell}
\usepackage{xcolor}
\usepackage{colortbl}
\usepackage{caption}
\usepackage{enumitem}
\usepackage{titlesec}
\usepackage[hidelinks]{hyperref}
\usepackage[expansion=false,protrusion=false]{microtype}
\usepackage{placeins}
\definecolor{ink}{HTML}{1B2A4A}
\definecolor{accent}{HTML}{C0392B}
\titleformat{\section}{\large\bfseries\color{ink}}{\thesection}{0.6em}{}
\titleformat{\subsection}{\normalsize\bfseries\color{ink}}{\thesubsection}{0.6em}{}
\newcommand{\SR}{\mathrm{SR}}
\newcommand{\PnL}{\mathrm{PnL}}
\title{\bfseries\color{ink}\Large Drawdown Risk Beyond Brownian Motion\\[3pt]
\large A Monte-Carlo Framework, Non-Gaussian Extensions, and Long Memory}
\author{Francesco Landolfi\\[3pt]
{\normalsize Epiphany, Imperia, Italy}\\[2pt]
{\small\texttt{francesco.landolfi@epiphany-alpha.com}}}
\date{July 2026}
\begin{document}
\maketitle
\begin{abstract}
\noindent How deep and how long should the drawdowns of a systematic trading strategy run, given its
Sharpe ratio and the statistical structure of its returns? Building on the drawdown framework of Rej,
Seager and Bouchaud~\cite{rsb2017}, we develop the answer in three steps. We first reframe their
closed-form results as a transparent Monte-Carlo experiment, validate it against their analytic
benchmarks, and extend the mapping from drawdowns to four decision-relevant measures: maximum drawdown,
maximum loss, final negative time and longest recovery time. We then relax the Gaussian
assumption, holding the true Sharpe and volatility fixed while varying skewness, fat tails, volatility
clustering and Sharpe-estimation uncertainty across strategy archetypes; the four measures move
differently, so a single Gaussian table mis-warns. We finally replace short-memory persistence
with fractional Brownian motion and show that the apparent amplification of drawdown risk under
persistence is, for maximum-drawdown depth, almost entirely self-similar dispersion scaling
($T^{H-1/2}$) rather than path geometry: a failure of square-root-of-time calibration, not intrinsic
danger. We provide reproducible lookup tables and a practical calibration recipe.\\[3pt]
\end{abstract}
\section{Introduction}
Every manager of a systematic strategy eventually faces the same decision. A live book is in a drawdown, and it must be judged either normal statistical pain, to be endured, or evidence that the edge has decayed and the strategy should be cut. The Sharpe ratio, the headline statistic used to select and rank strategies, says almost nothing about this. It is a single backward-looking number; it does not tell a manager how deep a drawdown can run by chance, how long a strategy can sit under water, or how much patience is required before performance is realised.

Rej, Seager and Bouchaud (RSB)~\cite{rsb2017} put numbers on exactly this question. Modelling the profit and loss (P\&L) as a drifted Brownian motion normalised to unit volatility, so that the annualised Sharpe ratio equals the drift, they derived closed-form distributions for the depth and length of a drawdown as functions of that single number, turning drawdown pain into a test of the Sharpe assumption.

This paper builds on their framework and develops it in three directions. First (Sections~\ref{sec:framework}--\ref{sec:mapping}), we reframe the RSB analysis as a transparent Monte-Carlo experiment, validate it against their analytic benchmarks, and extend the mapping from drawdowns to four measures that dominate the real-time keep-or-kill decision: maximum drawdown, maximum loss, final negative time and longest recovery time, cast as lookup tables a risk manager can read directly. Second (Section~\ref{sec:gaussian}), we relax the Gaussian assumption: holding the true Sharpe and volatility fixed, we vary skewness, fat tails, volatility clustering and Sharpe-estimation uncertainty across strategy archetypes, and show that the four measures do not move together, so a single Gaussian table systematically mis-warns. Third (Sections~\ref{sec:longmemory}--\ref{sec:synthesis}), we push persistence to its long-memory limit, fractional Brownian motion, and show that the apparent amplification of drawdown risk under persistence is, for maximum-drawdown depth, almost entirely a self-similar dispersion-scaling effect ($T^{H-1/2}$) rather than a deepening of path geometry: a failure of square-root-of-time calibration, not intrinsic danger.

The work sits in a line of quantitative research on the statistics of trading performance: on the option-like payoff of trend following~\cite{potters2005,lemperiere2014} and on the overfitting risk incurred when its signals are refined beyond what the data can identify~\cite{valeyre2025}, on the finite-sample uncertainty of the Sharpe ratio~\cite{lo2002}, on fractional Brownian motion and its occupation-time (arcsine) laws~\cite{mvn1968,sadhu2017}, and on the long memory of order flow that motivates persistence in strategy P\&L~\cite{bgpw2004,lillofarmer2004}. Every figure and table is reproducible, and Section~\ref{sec:practical} gives a practical recipe for calibrating the tables to a manager's own returns.

\section{The Monte-Carlo framework}\label{sec:framework}
\subsection{Modelling framework}
\label{p1:sec:model}
Following RSB we model the (log-)P\&L of a strategy as a Brownian motion with drift,
\begin{equation}
\mathrm{d}\,\PnL = \mu\,\mathrm{d}t + \sigma\,\mathrm{d}W ,
\end{equation}
over a finite horizon $(0,T)$. We normalise the risk to $\sigma = 1$, so that P\&L is measured in
\emph{units of annualised volatility}. With this normalisation the annualised Sharpe ratio is simply
the drift,
\begin{equation}
\SR = \frac{\mu}{\sigma} = \mu .
\end{equation}
Two consequences are worth stating explicitly, because they make every later chart dimensionless and
comparable across strategies: depth-type measures (drawdown, loss) are expressed as multiples of the
annual volatility, and time-type measures (time under water, recovery time) are expressed in years.
We take $250$ trading days per year.

\subsubsection{Generating paths with a prescribed Sharpe and volatility}
We model the P\&L directly as the drifted Brownian motion of RSB, discretised at the daily frequency.
From i.i.d.\ standard-normal innovations $r_i \sim \mathcal{N}(0,1)$, $i = 1,\dots,250H$, where $H$ is
the horizon in years, the daily returns are
\begin{equation}
r'_i \;=\; \frac{\Sigma}{\sqrt{250}}\left(\frac{S}{\sqrt{250}} \;+\; r_i\right),
\label{p1:eq:transform}
\end{equation}
so that $\mathbb{E}[r'_i] = \Sigma S/250$ and $\mathrm{std}(r'_i) = \Sigma/\sqrt{250}$. Annualising
gives a \emph{process} Sharpe ratio of exactly $S$ and an annual volatility of $\Sigma$; throughout we
set $\Sigma = 1$. These are properties of the data-generating process, not of any single realisation:
each simulated path has its own \emph{realised} Sharpe and its own terminal P\&L, both of which
fluctuate, exactly as a live track record would. A sample of equity lines is obtained by drawing $N$
independent paths of the form~\eqref{p1:eq:transform} and accumulating them.

This distinction matters for risk calibration. It is tempting to standardise the innovations to exact
in-sample mean and variance before applying~\eqref{p1:eq:transform}; doing so, however, pins the terminal
P\&L to a constant and turns the process into a Brownian \emph{bridge} --- artificially mean-reverting,
and one that systematically under-states drawdown depth and time under water, most severely at low
Sharpe, where the deepest and longest drawdowns are precisely the paths that end low. We therefore fix
only the process parameters and let the realised path statistics vary; this is both the correct object
for the question and what reproduces the RSB closed form, as the validation below confirms.\footnote{We
are grateful to readers of the first release who flagged the bias introduced by an earlier in-sample
standardisation of the innovations.}

\subsubsection{What does ``a Sharpe of $S$'' mean over a horizon?}
\begin{figure}[htbp]
\centering
\includegraphics[width=\linewidth]{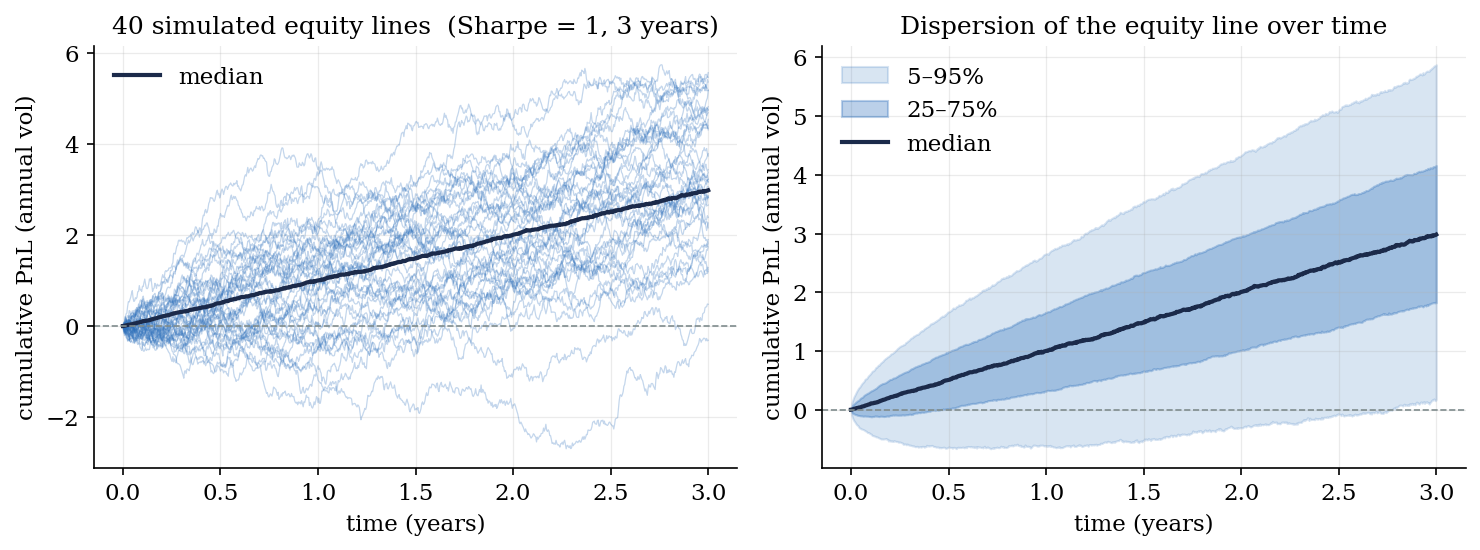}
\caption{Left: forty independent simulated equity lines for a strategy with Sharpe $=1$ over three
years ($\sigma=1$ normalisation, so the vertical axis is in units of annual volatility). Right: the
$5$--$95\%$ and $25$--$75\%$ dispersion bands of the same ensemble. Even for a respectable Sharpe of
one, the range of plausible equity levels remains large throughout the life of the strategy.}
\label{p1:fig:equity}
\end{figure}

A subtlety that the headline number hides is the \emph{implied horizon}. Saying ``my strategy has a
Sharpe of $S$'' leaves unsaid \emph{at what point in time} that figure is realised. If $S$ is the
Sharpe achieved over two years, then after one year the realised annual Sharpe can be almost
anything; only at the two-year mark is it exactly $S$. Figure~\ref{p1:fig:equity} makes this concrete:
forty simulated equity lines for $S=1$ over three years, and the corresponding dispersion fan. The
spread is wide for a long time, which is precisely why a single in-flight observation is such a noisy
signal of strategy quality.

\subsection{Four critical performance measures}
When monitoring a live strategy, a handful of risk measures carry disproportionate weight --- not
only analytically, but emotionally, because they govern the patience and conviction required to stay
the course. We focus on four.

\begin{description}[leftmargin=1.4em,style=nextline,itemsep=2pt]
\item[\textcolor{ink}{Maximum drawdown}] the largest peak-to-trough decline of the equity line over
the life of the strategy. It captures the depth of the worst loss \emph{relative to a prior high} and
is the classic measure of pain.
\item[\textcolor{ink}{Maximum loss}] the deepest absolute level the equity line reaches below its
starting point. Where drawdown is measured from the running peak, max loss is measured from zero; it
quantifies the worst-case absolute setback and the capital genuinely at risk.
\item[\textcolor{ink}{Final negative time}] the last point in time at which cumulative P\&L is still
negative. It measures how long the strategy can keep a manager waiting before it is unambiguously in
profit.
\item[\textcolor{ink}{Longest recovery time}] the longest stretch spent away from a previous equity
high --- the longest gap between successive record highs, sometimes called the time of absorption of
a drawdown. It measures the persistence required to claw back to break-even after a setback.
\end{description}

\begin{figure}[htbp]
\centering
\includegraphics[width=0.92\linewidth]{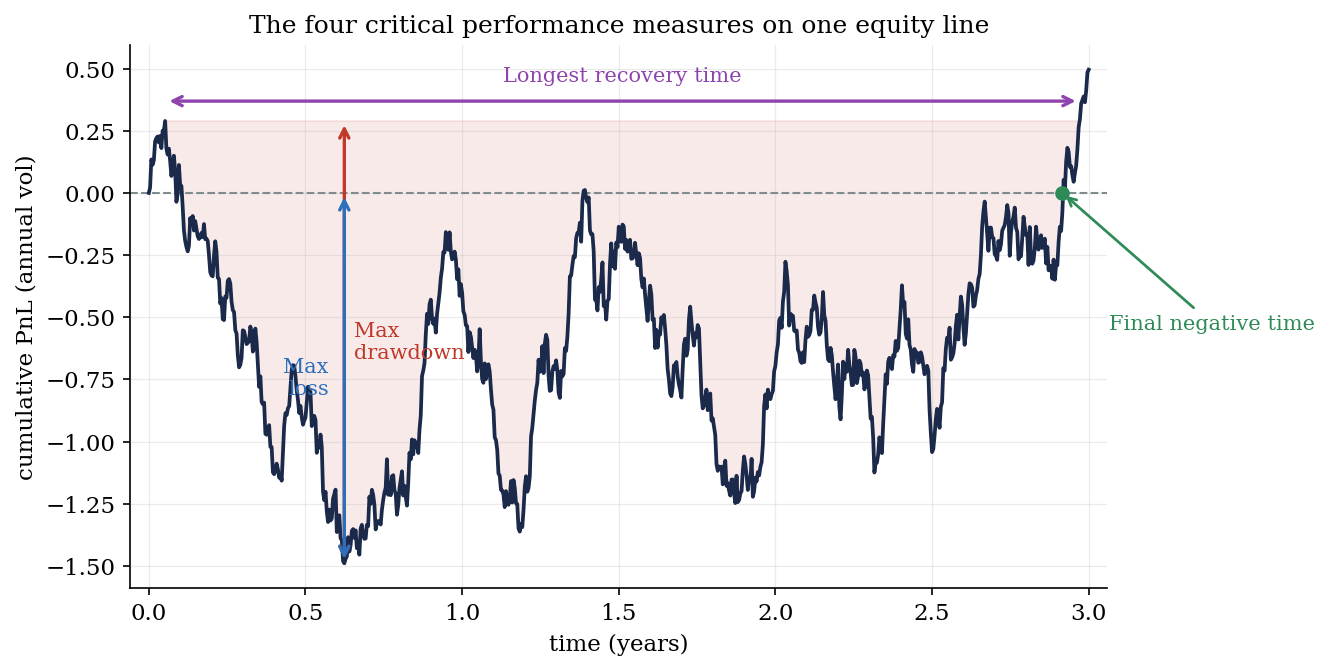}
\caption{The four critical performance measures on one simulated equity line. Maximum drawdown is the
largest drop from a running peak; maximum loss is the deepest excursion below the starting level;
final negative time is the last instant the cumulative P\&L is non-positive; longest recovery time is
the longest interval between successive all-time highs.}
\label{p1:fig:schematic}
\end{figure}

Figure~\ref{p1:fig:schematic} illustrates all four on a single simulated equity line.

Formally, let $X_t = \sum_{i\le t} r'_i$ denote the cumulative P\&L and
$M_t = \max_{s\le t} X_s$ its running maximum. The measures are
\begin{align}
\text{max drawdown} &= \max_{t\le T}\,(M_t - X_t), &
\text{max loss} &= \max\!\Big(0,\,-\min_{t\le T} X_t\Big), \\[2pt]
\text{final negative time} &= \max\{\,t \le T : X_t \le 0\,\}, &
\text{longest recovery} &= \max_k\,(\tau_{k+1}-\tau_k),
\end{align}
where $\{\tau_k\}$ are the times at which $X$ sets a new record high.

\subsection{Methodology: quantile analysis}
\begin{figure}[htbp]
\centering
\includegraphics[width=\linewidth]{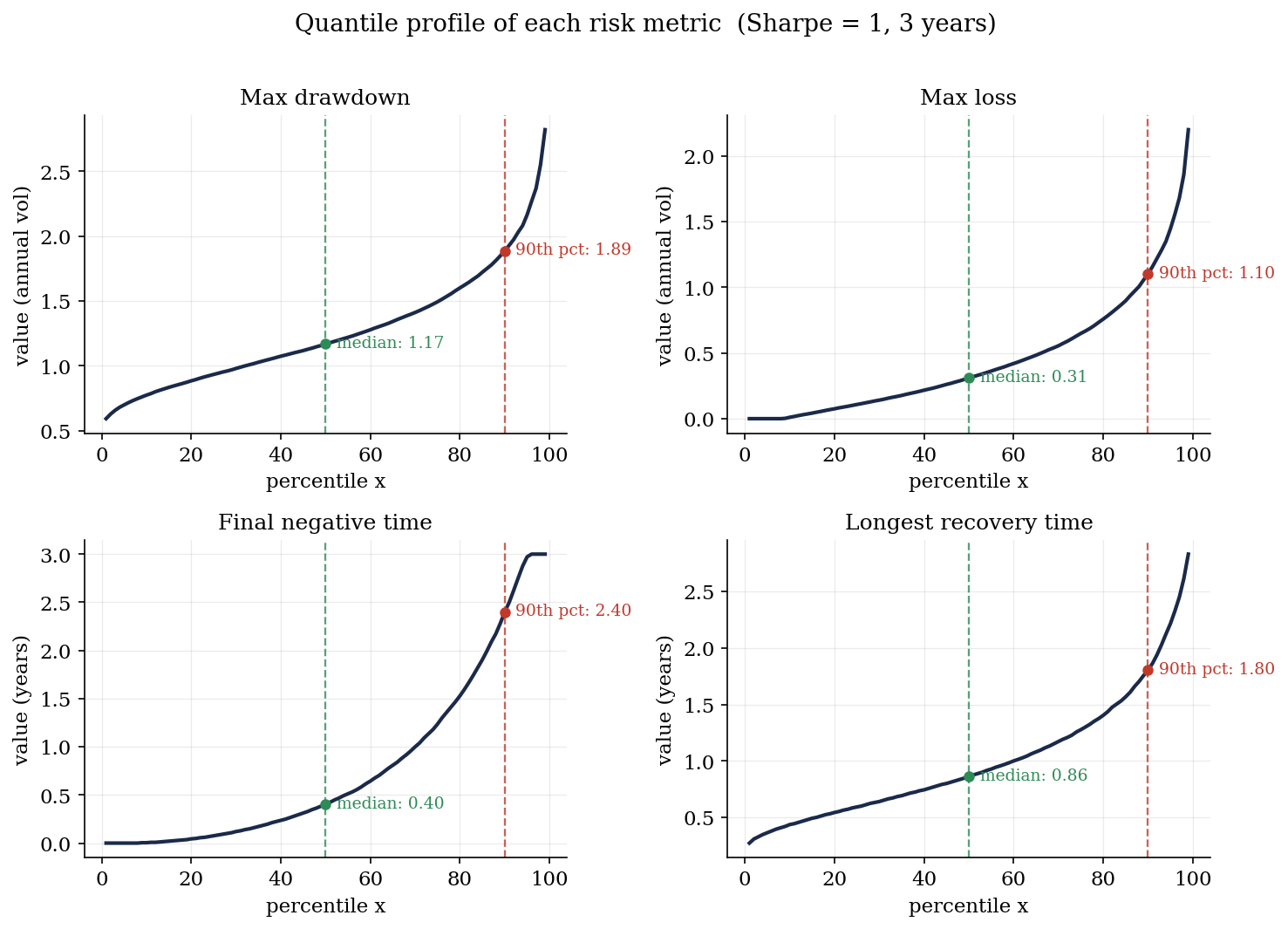}
\caption{Quantile profile of each risk measure for a strategy with Sharpe $=1$ over a three-year
horizon ($N=20{,}000$ simulated paths). The median (green) and the $90$th percentile (red) are
highlighted: these are the two reference levels carried into the maps and tables that follow.}
\label{p1:fig:quantiles}
\end{figure}

For a given assumption about Sharpe ratio, annual volatility and horizon, we draw a large sample of
equity lines and apply each metric to every path. This yields, for each measure, a vector of
realisations whose empirical distribution we summarise by quantiles. The quantile at level $x$ is the
threshold below which a fraction $x$ of the paths fall; reading the curve at $x=0.5$ gives the
\emph{expected} (median) case, while $x=0.9$ gives a conservative \emph{near-worst} case that occurs
only one time in ten. Figure~\ref{p1:fig:quantiles} shows the full quantile profile of each measure for
$S=1$ over three years.

\subsection{Validation against the analytic benchmark}
\begin{figure}[htbp]
\centering
\includegraphics[width=\linewidth]{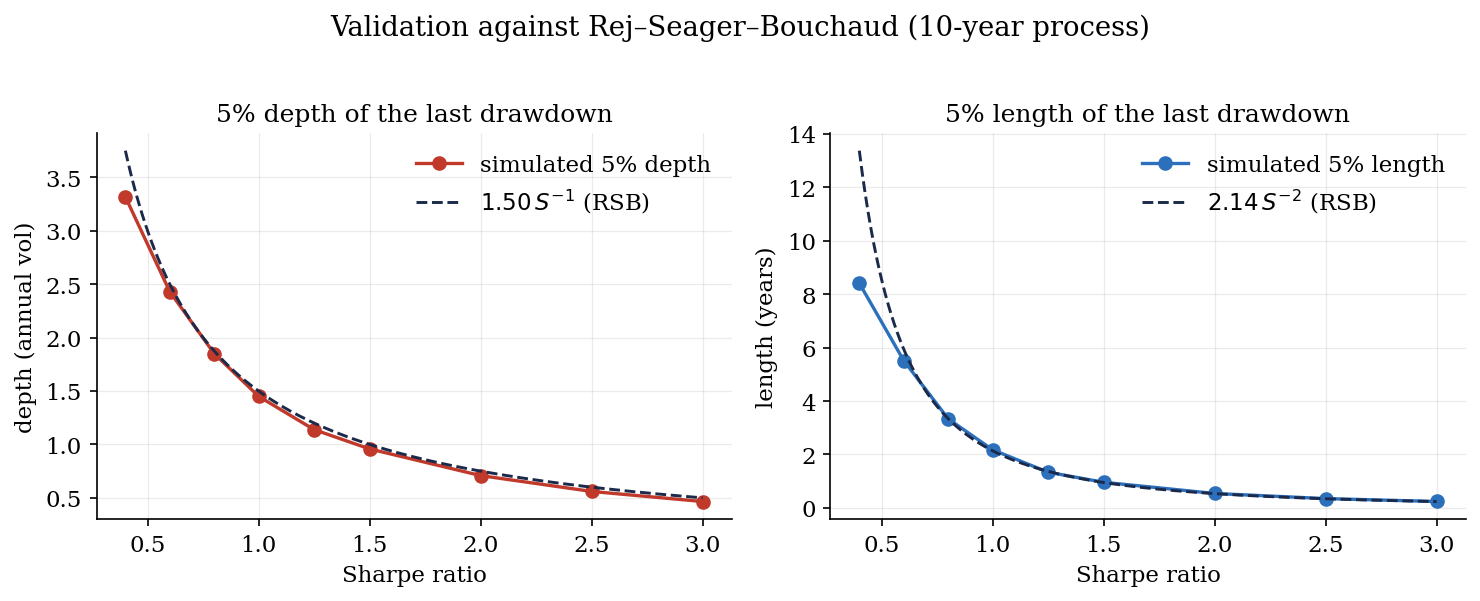}
\caption{Validation against Rej--Seager--Bouchaud for a ten-year process. The $5\%$ depth (left) and $5\%$
length (right) of the last drawdown from three sources: our numerical integration of the exact
densities $\rho(\ell)$ and $\psi(d)$ (solid green); the authors' published fits $1.50\,\SR^{-1}$ and
$2.14\,\SR^{-2}$ (dotted); and our Monte-Carlo estimate over $20{,}000$ paths (markers). The exact
density coincides with the published fit; the simulation tracks both, converging at higher Sharpe.}
\label{p1:fig:validation}
\end{figure}

Before extending the analysis, we check the engine against the closed-form results of RSB. For the
\emph{last} drawdown of a ten-year process they show that the depth and length consistent with the
Sharpe ratio at the $5\%$ level obey
\begin{equation}
d_{5\%} \approx 1.50\,\SR^{-1}, \qquad \ell_{5\%} \approx 2.14\,\SR^{-2},
\end{equation}
i.e.\ depth scales inversely with the Sharpe ratio and length inversely with its \emph{square}.
These follow from the unconditional densities of the length $\rho(\ell)$ and depth $\psi(d)$ of the
last drawdown,
\begin{align}
\rho(\ell) &= 2\!\left(\tfrac{1}{\sqrt{T-\ell}}\phi(\mu\sqrt{T-\ell})+\mu\,\Phi(\mu\sqrt{T-\ell})\right)
\!\left(\tfrac{1}{\sqrt{\ell}}\phi(\mu\sqrt{\ell})-\mu\,\Phi(-\mu\sqrt{\ell})\right), \\
\psi(d) &= \tfrac{d}{\pi}\,e^{-\mu d-\mu^2 T/2}\!\int_0^T\!\! \mathrm{d}\ell\,
\frac{e^{-d^2/2\ell}}{(\ell(T-\ell))^{3/2}}\!\left(T-\ell+\sqrt{2\pi}\,\mu(T-\ell)^{3/2}
e^{\mu^2(T-\ell)/2}\Phi(\mu\sqrt{T-\ell})\right),
\end{align}
with $\phi$ and $\Phi$ the standard-normal pdf and cdf. We validate at two levels. First, we
integrate the exact densities $\rho$ and $\psi$ numerically and extract their $5\%$ tail quantiles;
these reproduce the authors' published fits to three significant figures --- for example, at $\SR=1$
we obtain $\ell_{5\%}=2.144$ (fit $2.140$) and $d_{5\%}=1.498$ (fit $1.500$) --- confirming that our
implementation of the densities is correct. Second, we reproduce the authors' definitions in the
simulator --- depth $d=M_T-X_T$ and length $\ell = T - \arg\max_t X_t$ of the drawdown in progress at
$T$ --- and take the $95$th percentile across $20{,}000$ ten-year paths. Figure~\ref{p1:fig:validation}
overlays all three: the exact density sits essentially on top of the published fit, and the
Monte-Carlo estimate tracks both. With the process-level construction of Section~\ref{p1:sec:model} (no in-sample standardisation), the Monte-Carlo estimate now sits essentially on the exact density across the whole range. The only visible shortfall is at very low Sharpe, and it is benign: pure finite-horizon truncation --- a ten-year process simply cannot exhibit the thirteen-year drawdown the asymptotic law predicts at $\SR=0.4$. The qualitative laws --- depth
$\propto \SR^{-1}$, length $\propto \SR^{-2}$ --- are reproduced cleanly.

\section{Mapping the Sharpe ratio to the four measures}\label{sec:mapping}
\subsection{Results: mapping the Sharpe ratio to each measure}
\begin{figure}[htbp]
\centering
\includegraphics[width=\linewidth]{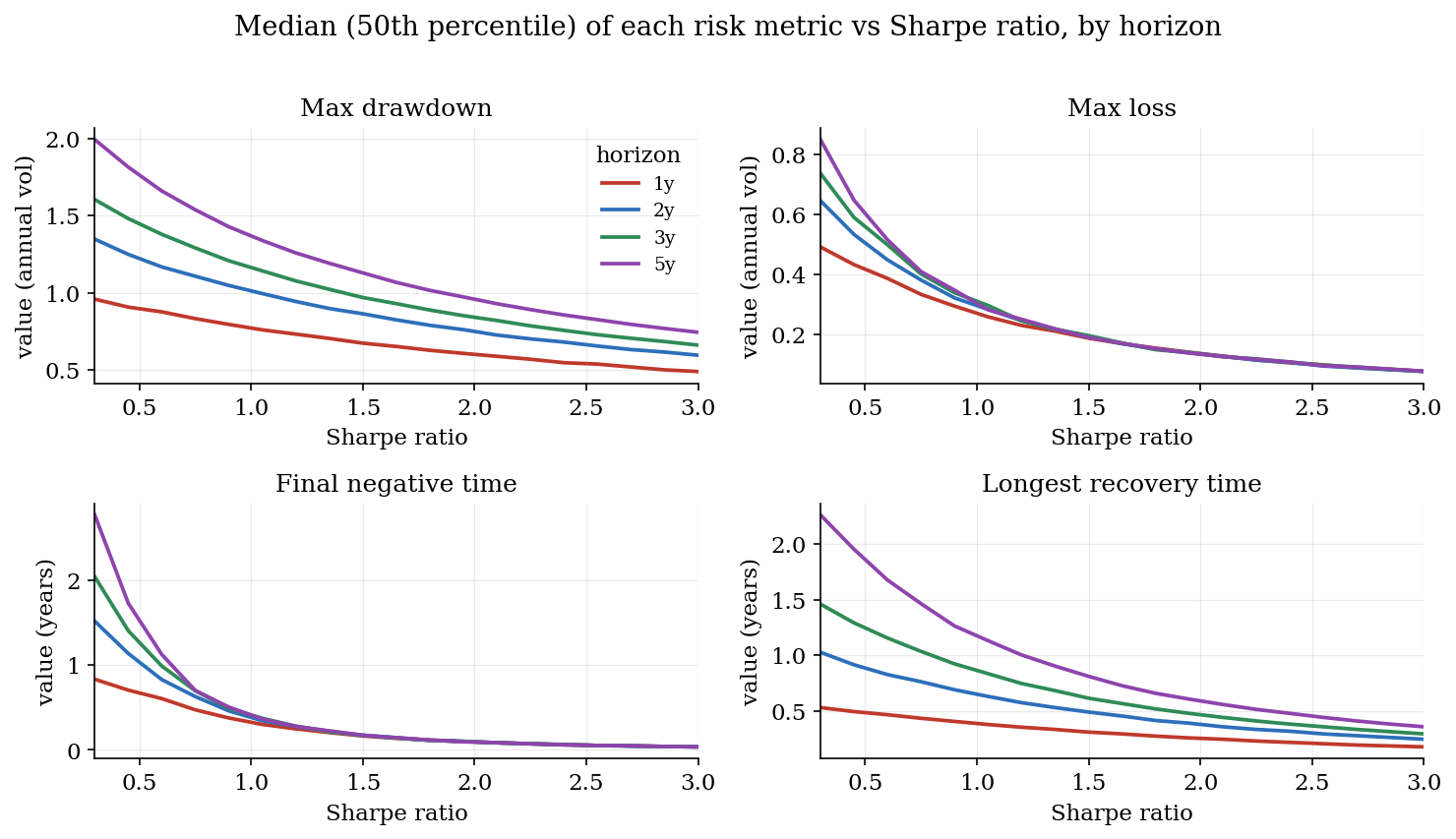}
\caption{Median (50th-percentile) value of each risk measure as a function of the Sharpe ratio, for
horizons of one, two, three and five years.}
\label{p1:fig:maps_median}
\end{figure}

\begin{figure}[htbp]
\centering
\includegraphics[width=\linewidth]{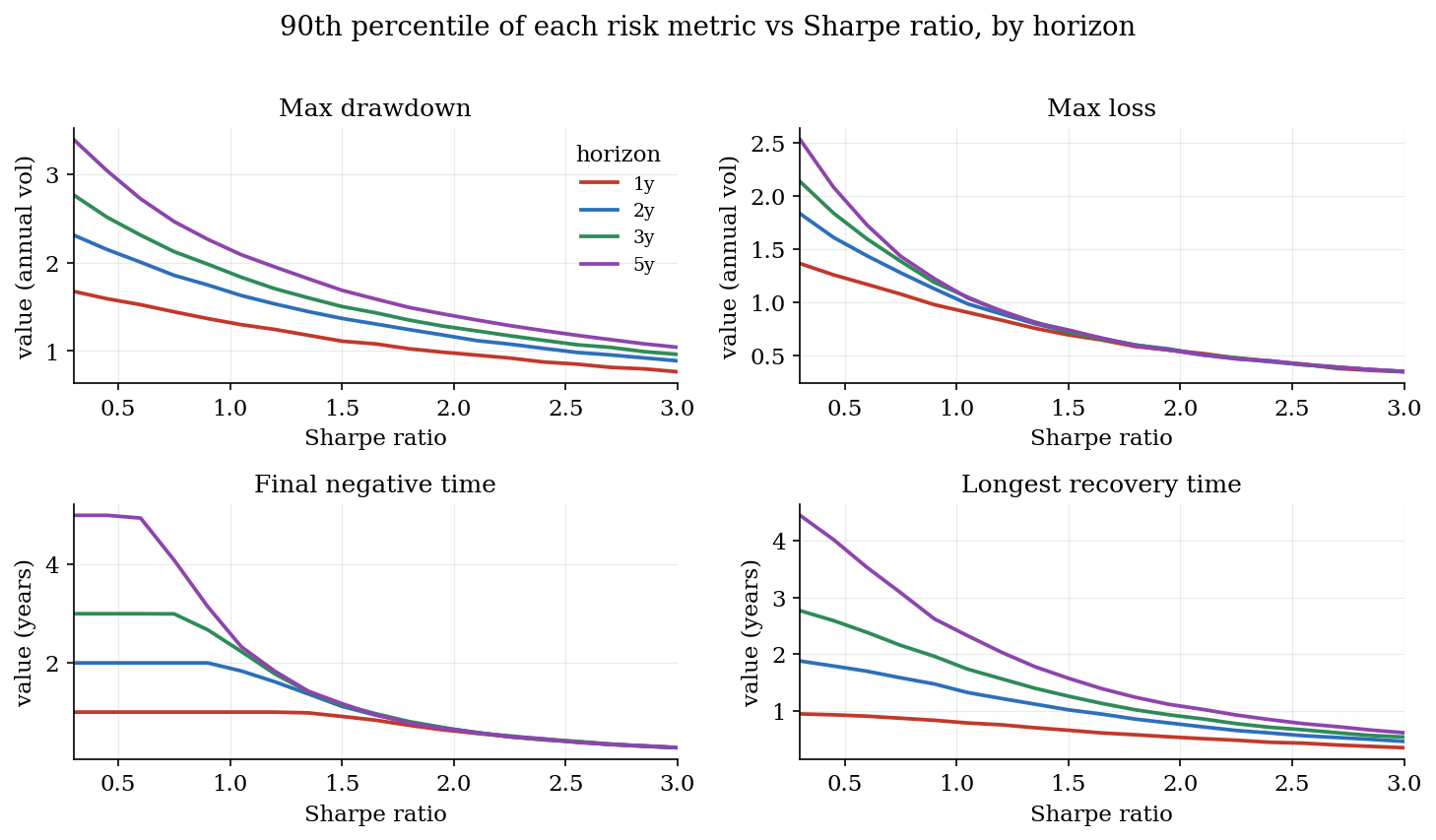}
\caption{$90$th-percentile (near-worst, one-in-ten) value of each risk measure as a function of the
Sharpe ratio, by horizon. These are the levels a manager should treat as warning thresholds rather
than expectations.}
\label{p1:fig:maps_p90}
\end{figure}

Having validated the machinery, we map all four measures against the Sharpe ratio for horizons of
one, two, three and five years. Figure~\ref{p1:fig:maps_median} shows the median (expected) outcome and
Figure~\ref{p1:fig:maps_p90} the $90$th-percentile (near-worst) outcome. Two patterns recur. Depth-type
measures (drawdown, loss, in units of annual vol) fall steadily with the Sharpe ratio and are only
weakly sensitive to horizon. Time-type measures (final negative time, recovery time) fall even faster
with Sharpe and grow markedly with horizon: the longer a strategy runs, the longer the worst dry
spell it will contain, even at a fixed Sharpe.

\FloatBarrier
\subsection{Decision tables}
The four maps condense into a single panel a manager --- or an allocator --- can keep beside a live
P\&L monitor (Table~\ref{p1:tab:panel}). It is built to be \emph{consulted}, not read: locate the assumed
Sharpe ratio and the elapsed horizon, and the cell tells you at a glance both the outcome to expect and
the threshold past which the drawdown stops being normal. The colour does the talking --- the redder the
cell, the more punishing the near-worst case for that measure.

\begin{table}[ht]\centering
\caption{\textbf{The drawdown decision panel.} Each cell shows the \emph{expected} case (left, the median) and the \emph{worry line} (right, the 90th percentile: a one-in-ten outcome under the Brownian benchmark), for a strategy of assumed Sharpe ratio (rows) and elapsed horizon (columns). Cells are shaded from calm green to alarm red by the severity of the worry line within each measure. \textbf{How to use it:} find your assumed Sharpe and how long the strategy has been live, then compare your current reading to the cell. Below the left number is business as usual; past the right number you are in tail territory --- the evidence-based moment to revise the Sharpe down or cut. Depth measures are in units of annual volatility; time measures in years.}
\label{p1:tab:panel}
\renewcommand{\arraystretch}{1.55}
\setlength{\tabcolsep}{9pt}
\small
\begin{tabular}{l|cccc}
\arrayrulecolor{gray!50}\hline
\rowcolor{ink}\textcolor{white}{\textbf{Sharpe}} & \textcolor{white}{\textbf{1 year}} & \textcolor{white}{\textbf{2 years}} & \textcolor{white}{\textbf{3 years}} & \textcolor{white}{\textbf{5 years}}\\
\hline
\multicolumn{5}{l}{\rule{0pt}{2.6ex}\textbf{\textcolor{ink}{Max drawdown}}~\textnormal{\footnotesize(annual vol)}}\\[1pt]
\;0.5 & \cellcolor[HTML]{EFF0CB}0.90\,$\rightarrow$\,\textbf{1.57} & \cellcolor[HTML]{FDE8C1}1.23\,$\rightarrow$\,\textbf{2.11} & \cellcolor[HTML]{FAD2B8}1.45\,$\rightarrow$\,\textbf{2.45} & \cellcolor[HTML]{F4B2AC}1.78\,$\rightarrow$\,\textbf{2.96}\\
\;1.0 & \cellcolor[HTML]{E4EFD0}0.77\,$\rightarrow$\,\textbf{1.32} & \cellcolor[HTML]{F3F0C9}1.01\,$\rightarrow$\,\textbf{1.68} & \cellcolor[HTML]{FCF1C5}1.16\,$\rightarrow$\,\textbf{1.89} & \cellcolor[HTML]{FDE5BF}1.37\,$\rightarrow$\,\textbf{2.16}\\
\;1.5 & \cellcolor[HTML]{DDEED3}0.68\,$\rightarrow$\,\textbf{1.13} & \cellcolor[HTML]{E7EFCE}0.86\,$\rightarrow$\,\textbf{1.39} & \cellcolor[HTML]{ECF0CC}0.97\,$\rightarrow$\,\textbf{1.51} & \cellcolor[HTML]{F4F0C9}1.12\,$\rightarrow$\,\textbf{1.70}\\
\;2.0 & \cellcolor[HTML]{D6EED6}0.60\,$\rightarrow$\,\textbf{0.97} & \cellcolor[HTML]{DDEFD3}0.75\,$\rightarrow$\,\textbf{1.15} & \cellcolor[HTML]{E2EFD1}0.84\,$\rightarrow$\,\textbf{1.26} & \cellcolor[HTML]{E8EFCE}0.96\,$\rightarrow$\,\textbf{1.40}\\
\hline
\multicolumn{5}{l}{\rule{0pt}{2.6ex}\textbf{\textcolor{ink}{Max loss}}~\textnormal{\footnotesize(annual vol)}}\\[1pt]
\;0.5 & \cellcolor[HTML]{FDF1C5}0.42\,$\rightarrow$\,\textbf{1.23} & \cellcolor[HTML]{FBD8BA}0.52\,$\rightarrow$\,\textbf{1.57} & \cellcolor[HTML]{F8CBB5}0.56\,$\rightarrow$\,\textbf{1.73} & \cellcolor[HTML]{F4B2AC}0.61\,$\rightarrow$\,\textbf{2.02}\\
\;1.0 & \cellcolor[HTML]{ECF0CC}0.27\,$\rightarrow$\,\textbf{0.93} & \cellcolor[HTML]{F3F0C9}0.29\,$\rightarrow$\,\textbf{1.05} & \cellcolor[HTML]{F5F0C8}0.31\,$\rightarrow$\,\textbf{1.10} & \cellcolor[HTML]{F6F0C8}0.31\,$\rightarrow$\,\textbf{1.12}\\
\;1.5 & \cellcolor[HTML]{DFEFD2}0.20\,$\rightarrow$\,\textbf{0.69} & \cellcolor[HTML]{E1EFD1}0.19\,$\rightarrow$\,\textbf{0.73} & \cellcolor[HTML]{DFEFD2}0.19\,$\rightarrow$\,\textbf{0.70} & \cellcolor[HTML]{E1EFD1}0.19\,$\rightarrow$\,\textbf{0.73}\\
\;2.0 & \cellcolor[HTML]{D7EED6}0.14\,$\rightarrow$\,\textbf{0.54} & \cellcolor[HTML]{D7EED6}0.14\,$\rightarrow$\,\textbf{0.54} & \cellcolor[HTML]{D7EED6}0.14\,$\rightarrow$\,\textbf{0.54} & \cellcolor[HTML]{D6EED6}0.13\,$\rightarrow$\,\textbf{0.53}\\
\hline
\multicolumn{5}{l}{\rule{0pt}{2.6ex}\textbf{\textcolor{ink}{Final negative time}}~\textnormal{\footnotesize(years)}}\\[1pt]
\;0.5 & \cellcolor[HTML]{DDEFD3}0.67\,$\rightarrow$\,\textbf{1.00} & \cellcolor[HTML]{F0F0CB}1.05\,$\rightarrow$\,\textbf{2.00} & \cellcolor[HTML]{FEEBC2}1.29\,$\rightarrow$\,\textbf{3.00} & \cellcolor[HTML]{F4B2AC}1.55\,$\rightarrow$\,\textbf{4.99}\\
\;1.0 & \cellcolor[HTML]{DDEFD3}0.34\,$\rightarrow$\,\textbf{1.00} & \cellcolor[HTML]{EEF0CB}0.39\,$\rightarrow$\,\textbf{1.92} & \cellcolor[HTML]{F7F0C8}0.42\,$\rightarrow$\,\textbf{2.37} & \cellcolor[HTML]{FCF1C5}0.43\,$\rightarrow$\,\textbf{2.64}\\
\;1.5 & \cellcolor[HTML]{DCEED3}0.17\,$\rightarrow$\,\textbf{0.93} & \cellcolor[HTML]{E0EFD2}0.17\,$\rightarrow$\,\textbf{1.13} & \cellcolor[HTML]{DFEFD2}0.17\,$\rightarrow$\,\textbf{1.11} & \cellcolor[HTML]{E0EFD2}0.17\,$\rightarrow$\,\textbf{1.15}\\
\;2.0 & \cellcolor[HTML]{D6EED6}0.09\,$\rightarrow$\,\textbf{0.62} & \cellcolor[HTML]{D6EED6}0.09\,$\rightarrow$\,\textbf{0.64} & \cellcolor[HTML]{D7EED6}0.09\,$\rightarrow$\,\textbf{0.65} & \cellcolor[HTML]{D7EED6}0.09\,$\rightarrow$\,\textbf{0.65}\\
\hline
\multicolumn{5}{l}{\rule{0pt}{2.6ex}\textbf{\textcolor{ink}{Longest recovery time}}~\textnormal{\footnotesize(years)}}\\[1pt]
\;0.5 & \cellcolor[HTML]{DFEFD2}0.49\,$\rightarrow$\,\textbf{0.92} & \cellcolor[HTML]{F4F0C9}0.89\,$\rightarrow$\,\textbf{1.76} & \cellcolor[HTML]{FDE6C0}1.25\,$\rightarrow$\,\textbf{2.51} & \cellcolor[HTML]{F4B2AC}1.87\,$\rightarrow$\,\textbf{3.91}\\
\;1.0 & \cellcolor[HTML]{DDEFD3}0.39\,$\rightarrow$\,\textbf{0.82} & \cellcolor[HTML]{EAEFCD}0.65\,$\rightarrow$\,\textbf{1.37} & \cellcolor[HTML]{F5F0C8}0.85\,$\rightarrow$\,\textbf{1.80} & \cellcolor[HTML]{FEE9C1}1.17\,$\rightarrow$\,\textbf{2.43}\\
\;1.5 & \cellcolor[HTML]{D9EED5}0.32\,$\rightarrow$\,\textbf{0.67} & \cellcolor[HTML]{E2EFD1}0.48\,$\rightarrow$\,\textbf{1.02} & \cellcolor[HTML]{E7EFCE}0.62\,$\rightarrow$\,\textbf{1.25} & \cellcolor[HTML]{EFF0CB}0.80\,$\rightarrow$\,\textbf{1.58}\\
\;2.0 & \cellcolor[HTML]{D6EED6}0.26\,$\rightarrow$\,\textbf{0.53} & \cellcolor[HTML]{DCEED4}0.39\,$\rightarrow$\,\textbf{0.76} & \cellcolor[HTML]{DFEFD2}0.47\,$\rightarrow$\,\textbf{0.91} & \cellcolor[HTML]{E4EFD0}0.59\,$\rightarrow$\,\textbf{1.09}\\
\hline
\end{tabular}

\vspace{4pt}
{\footnotesize\centering
\colorbox[HTML]{D6EED6}{\strut~calm~} \;$\rightarrow$\; \colorbox[HTML]{FFF1C4}{\strut~watch~} \;$\rightarrow$\; \colorbox[HTML]{F4B2AC}{\strut~alarm~}\hspace{1.5em}\textbf{left} = expected (median)\quad\textbf{right} = worry line (90th pct, 1-in-10)\par}
\end{table}

\subsubsection{A worked decision rule}
Suppose a strategy is expected to deliver a Sharpe of $2$ over a two-year horizon. From
Table~\ref{p1:tab:panel}, a maximum drawdown of about $1.05\times$ the annual volatility already sits at
the $90$th percentile --- a one-in-ten event under the Brownian benchmark. If the live drawdown
exceeds that level, the manager faces a choice consistent with the RSB framing: either accept it as
bad luck (a genuine $10\%$ tail event), or revise the assumed Sharpe downward. If, in addition, the
drawdown is accompanied by an unusually long final-negative-time or recovery-time relative to the
panel, the case for a downward revision --- or for halting the strategy --- strengthens. The panel
turns an intuitive, emotional call into an evidence-based one.

\FloatBarrier

\section{Beyond the Gaussian: non-Gaussian return archetypes}\label{sec:gaussian}
Real strategy returns are not Gaussian, and the violations are characteristic of the strategy's style rather than random. This section relaxes the Brownian benchmark of Sections~\ref{sec:framework}--\ref{sec:mapping}: we hold the true Sharpe ratio and volatility fixed and vary only the higher-order structure of returns.

\subsection{What makes real equity lines non-Brownian}
We group the departures from the RSB benchmark into four, matched to the four knobs of our model.

\paragraph{Skewness.} The sign of skew is largely set by the strategy's payoff convexity. Convex,
option-like strategies (trend, breakout, long-volatility) are positively skewed: many small losses,
occasional large gains, and a win rate that can fall well below 50\% while remaining
profitable~\cite{potters2005}. Concave strategies (mean-reversion, carry, short-volatility) are
negatively skewed: a high win rate punctuated by rare, severe losses.

\paragraph{Fat tails.} Daily strategy returns have excess kurtosis far above the Gaussian, deepening
the worst-case depth measures regardless of skew sign~\cite{bouchaud_potters_book}.

\paragraph{Volatility clustering.} Volatility is not constant: losses arrive in bursts. Clustering
deepens and prolongs drawdowns relative to an i.i.d.\ series with the same marginal distribution, and
is especially pronounced for short-volatility-like profiles.

\paragraph{Persistence.} Strategy P\&L can be positively autocorrelated --- through slow or
overlapping positions, momentum in the strategy's own returns, or crowding. Even mild positive
autocorrelation lengthens time-under-water substantially; the analytic limit of this effect is
fractional Brownian motion, whose drawdown statistics obey generalised arcsine laws~\cite{sadhu2017}.

\paragraph{Sharpe uncertainty.} Finally, the headline Sharpe is an estimate. For an annualised Sharpe
$\hat S$ measured over $T_{\text{obs}}$ years of daily data, its standard error is approximately
$\sigma_S \approx \sqrt{(1+\hat S^2/2)\,/\,(250\,T_{\text{obs}})}\times\sqrt{250}
= \sqrt{(1+\hat S^2/2)/T_{\text{obs}}}$~\cite{lo2002} --- we use the conservative annual-frequency form.
For $\hat S=1$ over three years this is $\approx 0.71$: the true Sharpe could plausibly be anywhere from
$0.3$ to $1.7$. Marginalising over this uncertainty widens every drawdown distribution.

\subsection{The model}
We model daily returns as an AR(1)--GARCH(1,1) process driven by skewed, fat-tailed innovations.
Let $z_t$ be i.i.d.\ draws from a normal-inverse-Gaussian distribution with shape $a$ and asymmetry
$b$, standardised to zero mean and unit variance; the sign of $b$ controls skew and $a$ controls tail
weight. Volatility clustering and persistence are added through
\begin{align}
h_t &= \omega + \alpha\,\varepsilon_{t-1}^2 + \beta\,h_{t-1}, \qquad \varepsilon_t = \sqrt{h_t}\,z_t,
\qquad \omega = 1-\alpha-\beta, \\
y_t &= \phi\,y_{t-1} + \varepsilon_t .
\end{align}
The noise $y_t$ is normalised to zero mean and unit variance using \emph{pooled}, process-level
statistics --- $\tilde y_t = (y_t-\bar y)/s_y$, with $\bar y$ and $s_y$ taken across the whole
ensemble rather than per path --- and the process annualised Sharpe $S$ and volatility $\Sigma$ are
imposed by
\begin{equation}
r_t \;=\; \frac{\Sigma}{\sqrt{250}}\left(\frac{S_{\text{eff}}}{\sqrt{250}} + \tilde y_t\right),
\qquad \Sigma = 1 .
\label{p2:eq:transform2}
\end{equation}
Because the centring and scaling are global constants, not per-path operations, the \emph{process}
Sharpe and volatility are fixed identically across archetypes --- the headline numbers a manager would
quote --- while each path's \emph{realised} Sharpe and terminal P\&L stay free to vary, exactly as a
live track record would.\footnote{Normalising each path to its \emph{own} exact in-sample mean and
variance, as an earlier version did, instead pins the terminal P\&L and turns the process into a
Brownian bridge, biasing the drawdown statistics downward; we are grateful to readers who flagged this.
The construction here avoids it.} What differs across archetypes is only the shape and dynamics that
drive drawdowns. This is what makes the comparison fair.

To represent Sharpe uncertainty we set, for each simulated path, $S_{\text{eff}} \sim \mathcal{N}(S,\sigma_S)$
with $\sigma_S$ the Lo standard error above; the manager believes the Sharpe is $S$, but each realised
life of the strategy may be governed by a different true value. Equivalently, we read the Lo standard
error as the spread of plausible \emph{true} Sharpes around the reported one --- a Bayesian
interpretation under a flat prior. We deliberately leave this normal untruncated: at a low estimated
Sharpe it places non-trivial mass on a genuinely negative true Sharpe (for $S=0.5$ over three years,
about one chance in five), which is an honest part of the uncertainty penalty rather than a modelling
artefact.

\subsubsection{Four archetypes}
We define four stylised archetypes, with parameters informed by the literature rather than fitted to a
specific fund (Table~\ref{p2:tab:calib}). This is deliberate: as Valeyre~\cite{valeyre2025} argues in
the trend-following context, the data available are rarely sufficient to identify richer
specifications without a serious risk of overfitting, so a transparent stylised parameterisation is
the more honest object here. The ``Uncertain Sharpe'' scenario uses Gaussian structure with
$S_{\text{eff}}$ randomised as above. Each archetype varies several features at once --- skew, tails,
clustering and persistence move together, as they do in real strategies --- so what follows is a
comparison between \emph{styles} as they actually occur, not a one-factor attribution of risk to any
single moment.

\begin{table}[htbp]\centering
\caption{Archetype parameters and the resulting realised daily moments (Sharpe $=1$, three years,
pooled across paths). All four share the same \emph{process} Sharpe ($1.00$) and annual volatility
($1.00$) by construction; realised per-path values fluctuate around them.}
\label{p2:tab:calib}
\small\setlength{\tabcolsep}{4.2pt}
\begin{tabular}{lcccccccc}
\toprule
\rowcolor{ink!12}\textbf{Archetype} & \textbf{NIG} $a$ & \textbf{NIG} $b$ & \textbf{AR} $\phi$ & \textbf{GARCH} $\alpha$ & \textbf{GARCH} $\beta$ & \textbf{skew} & \textbf{exc.\ kurt.} & \textbf{lag-1 AC} \\
\midrule
Gaussian (RSB) & --- & --- & 0.00 & 0.00 & 0.00 & $0.00$ & $0.0$ & $0.00$ \\
Trend / breakout & 1.2 & $+0.7$ & 0.15 & 0.06 & 0.90 & $+1.74$ & $7.8$ & $0.13$ \\
Mean-reversion / short-vol & 1.2 & $-0.7$ & 0.00 & 0.12 & 0.86 & $-1.85$ & $11.3$ & $0.01$ \\
Market-neutral & 1.5 & $0.0$ & 0.03 & 0.05 & 0.90 & $0.00$ & $2.2$ & $0.04$ \\
\bottomrule
\end{tabular}
\end{table}

\paragraph{Simulation and reproducibility.} All quantiles are estimated from $16{,}000$ independent
Monte-Carlo paths per cell, each a $750$-day (three-year) equity line at $250$ trading days per year.
NIG innovations are drawn and standardised to zero mean and unit variance; the GARCH recursion is
initialised at its unconditional variance ($h_0=1$, since $\omega=1-\alpha-\beta$), so no burn-in is
required. The pooled mean and standard deviation used in the normalisation are computed across the full
ensemble: this is a Monte-Carlo construction that fixes the \emph{process} moments, not a quantity a
manager could observe on a single live track. Across the Sharpe grid (Figure~\ref{p2:fig:grid}) we use
\emph{common random numbers} --- the same simulated noise at every Sharpe, with only the drift changing
--- so the curves are smooth and their ordering reflects the model, not sampling noise. The random seed
is fixed for reproducibility.

\subsection{Results}
\subsubsection{Same true Sharpe, different equity lines}
\begin{figure}[htbp]\centering
\includegraphics[width=\linewidth]{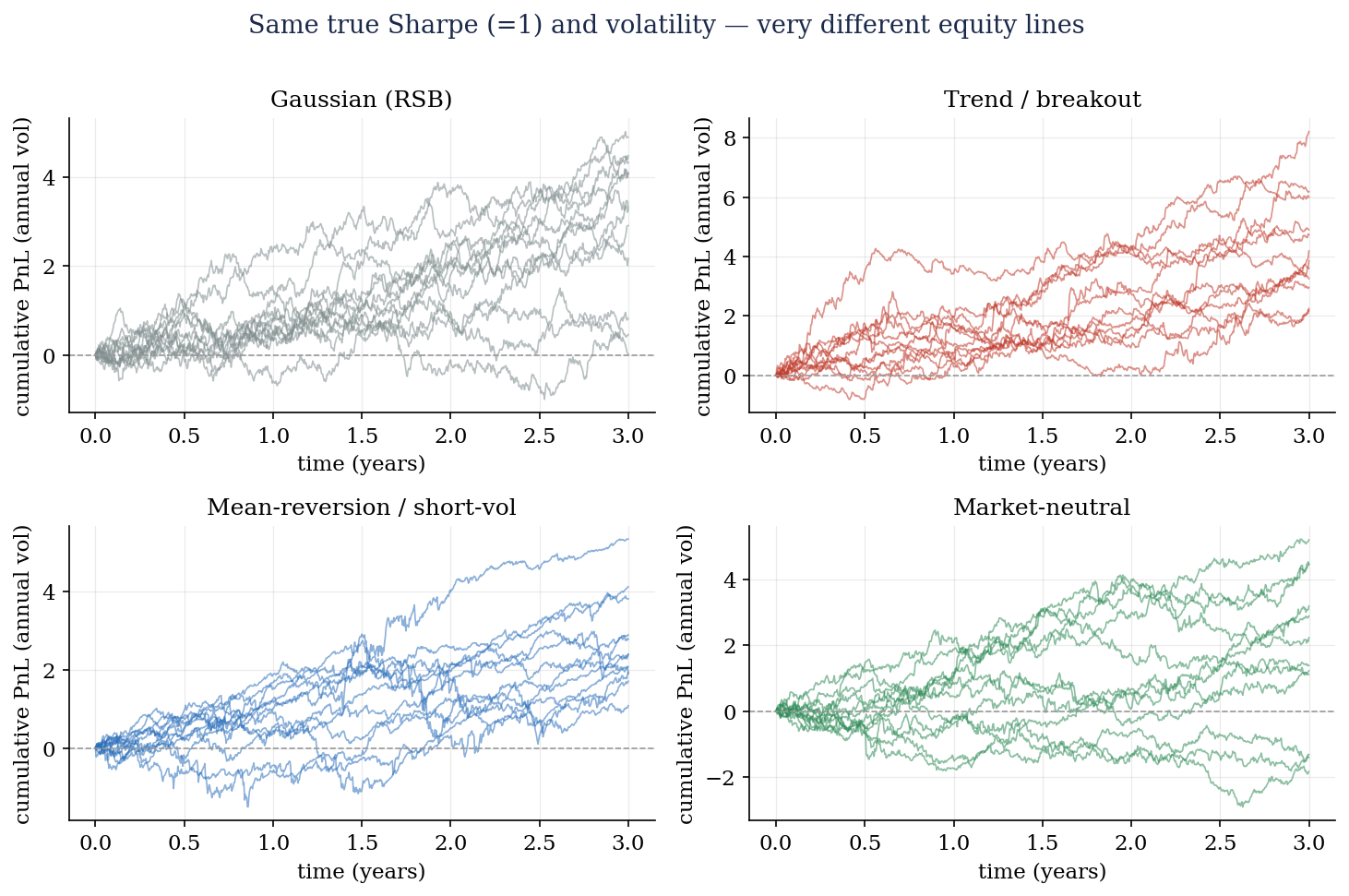}
\caption{Twelve independent three-year realisations per archetype, all drawn from a process with the same
true Sharpe ($=1$) and volatility; only the higher-order structure (skew, fat tails, volatility
clustering, persistence) differs across panels. Realised outcomes vary widely --- at a true Sharpe of
$1$, some three-year lives still end below zero. That is the point: the headline Sharpe fixes neither
the path nor its drawdowns.}
\label{p2:fig:paths}
\end{figure}

Figure~\ref{p2:fig:paths} shows twelve simulated three-year equity lines for each archetype, all drawn
from a process with a true Sharpe of $1$ and unit volatility. The Gaussian paths are the familiar
noisy upward drift; the trend paths grind sideways or down for long stretches and then jump; the
short-volatility paths climb smoothly and then suffer sharp, deep cliffs; the neutral paths sit close to
the Gaussian. The true Sharpe and volatility are identical across panels --- the realised paths, and
their drawdowns, are not.

\subsubsection{The depth--duration asymmetry}
\begin{figure}[htbp]\centering
\includegraphics[width=\linewidth]{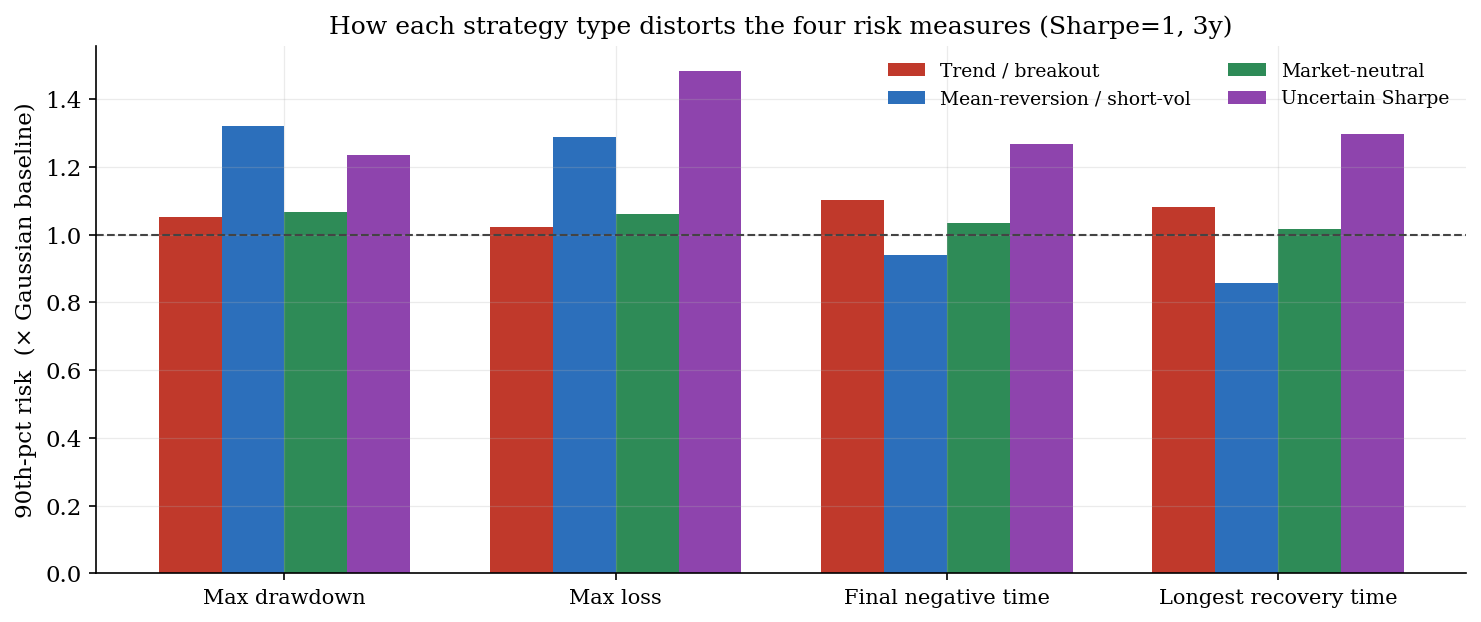}
\caption{The 90th-percentile of each risk measure, by archetype, relative to the Gaussian (RSB)
baseline (dashed line at $1$). The mean-reversion / short-vol book inflates the \emph{depth} measures
(max drawdown, max loss) by about a third while leaving the \emph{time} measures unchanged or lower;
the trend book instead lengthens the \emph{time} measures, with depth left near the benchmark;
uncertainty about the Sharpe lifts every measure at once, most of all maximum loss.}
\label{p2:fig:bars}
\end{figure}

The central result is that the four risk measures do \emph{not} move together; the strategy's higher-order
structure determines which becomes the binding constraint. Table~\ref{p2:tab:shift} and Figure~\ref{p2:fig:bars}
quantify this at Sharpe $1$ over three years.

\begin{table}[ht]\centering
\caption{\textbf{The style distortion map.} Every archetype carries the \emph{same} process Sharpe ($=1$) and volatility over a three-year horizon, yet the four risk measures diverge. Each cell shows median\,/\,90th-percentile; the figure below is the 90th-percentile relative to the Gaussian (RSB) baseline. Cells are shaded by that multiple --- \colorbox[HTML]{CFE8CF}{\strut\,green\,} where the measure is milder than the Gaussian, \colorbox[HTML]{F4B2AC}{\strut\,red\,} where it is worse. Depth measures in units of annual volatility; time measures in years.}
\label{p2:tab:shift}
\renewcommand{\arraystretch}{1.7}\setlength{\tabcolsep}{8pt}\small
\begin{tabular}{l|cccc}
\arrayrulecolor{gray!50}\hline
\rowcolor{ink}\textcolor{white}{\textbf{~Archetype}} & \textcolor{white}{\textbf{Max drawdown}} & \textcolor{white}{\textbf{Max loss}} & \textcolor{white}{\textbf{Final negative time}} & \textcolor{white}{\textbf{Longest recovery time}}\\
\hline
~Gaussian (RSB) & \cellcolor{gray!12}\makecell{1.16\,/\,1.88\\[-1pt]\scriptsize baseline} & \cellcolor{gray!12}\makecell{0.31\,/\,1.09\\[-1pt]\scriptsize baseline} & \cellcolor{gray!12}\makecell{0.42\,/\,2.37\\[-1pt]\scriptsize baseline} & \cellcolor{gray!12}\makecell{0.86\,/\,1.82\\[-1pt]\scriptsize baseline}\\
~Trend / breakout & \cellcolor[HTML]{F4EFF0}\makecell{1.24\,/\,\textbf{1.97}\\[-1pt]\scriptsize(1.05$\times$)} & \cellcolor[HTML]{F4F2F4}\makecell{0.32\,/\,\textbf{1.12}\\[-1pt]\scriptsize(1.02$\times$)} & \cellcolor[HTML]{F4E9E9}\makecell{0.49\,/\,\textbf{2.61}\\[-1pt]\scriptsize(1.10$\times$)} & \cellcolor[HTML]{F4EBEC}\makecell{0.98\,/\,\textbf{1.96}\\[-1pt]\scriptsize(1.08$\times$)}\\
~Mean-reversion / short-vol & \cellcolor[HTML]{F4CECB}\makecell{1.05\,/\,\textbf{2.48}\\[-1pt]\scriptsize(1.32$\times$)} & \cellcolor[HTML]{F4D2D0}\makecell{0.22\,/\,\textbf{1.41}\\[-1pt]\scriptsize(1.29$\times$)} & \cellcolor[HTML]{EBF3ED}\makecell{0.25\,/\,\textbf{2.23}\\[-1pt]\scriptsize(0.94$\times$)} & \cellcolor[HTML]{DFF0DF}\makecell{0.62\,/\,\textbf{1.56}\\[-1pt]\scriptsize(0.86$\times$)}\\
~Market-neutral & \cellcolor[HTML]{F4EDEE}\makecell{1.21\,/\,\textbf{2.00}\\[-1pt]\scriptsize(1.07$\times$)} & \cellcolor[HTML]{F4EEEF}\makecell{0.30\,/\,\textbf{1.16}\\[-1pt]\scriptsize(1.06$\times$)} & \cellcolor[HTML]{F4F1F2}\makecell{0.42\,/\,\textbf{2.45}\\[-1pt]\scriptsize(1.03$\times$)} & \cellcolor[HTML]{F4F3F5}\makecell{0.87\,/\,\textbf{1.85}\\[-1pt]\scriptsize(1.02$\times$)}\\
~Uncertain Sharpe & \cellcolor[HTML]{F4D8D7}\makecell{1.18\,/\,\textbf{2.32}\\[-1pt]\scriptsize(1.23$\times$)} & \cellcolor[HTML]{F4BAB5}\makecell{0.33\,/\,\textbf{1.62}\\[-1pt]\scriptsize(1.48$\times$)} & \cellcolor[HTML]{F4D4D3}\makecell{0.44\,/\,\textbf{3.00}\\[-1pt]\scriptsize(1.27$\times$)} & \cellcolor[HTML]{F4D1CE}\makecell{0.87\,/\,\textbf{2.36}\\[-1pt]\scriptsize(1.30$\times$)}\\
\hline
\end{tabular}
\end{table}

Three readings stand out. \emph{The short-volatility book is a depth problem}: its near-worst maximum
drawdown and maximum loss rise to roughly $1.32\times$ and $1.29\times$ the Gaussian benchmark, yet its
time-under-water is, if anything, slightly shorter --- it falls hard but climbs back quickly.
\emph{The trend book is a duration problem}: its largest distortion is in recovery time and
time-under-water ($1.08$--$1.10\times$), consistent with long grinding droughts between rare winning
trades; notably its depth stays essentially at the benchmark ($1.05\times$), because the positive skew
that should have made its drawdowns shallower is offset by its fat tails and volatility clustering ---
there is no depth dividend, only a shift of the binding constraint toward duration.
\emph{Market-neutral} stays within about $7\%$ of the Gaussian on every measure, vindicating the RSB
benchmark for genuinely diversified, symmetric books. The distortions are real but asymmetric in size:
the depth penalty on a short-volatility book ($\sim+30\%$) is far larger than the duration penalty on a
trend book ($\sim+10\%$). A single Gaussian table therefore under-warns a short-volatility manager on
depth and, more mildly, a trend manager on duration.

\subsubsection{Scaling with the Sharpe ratio}
\begin{figure}[htbp]\centering
\includegraphics[width=\linewidth]{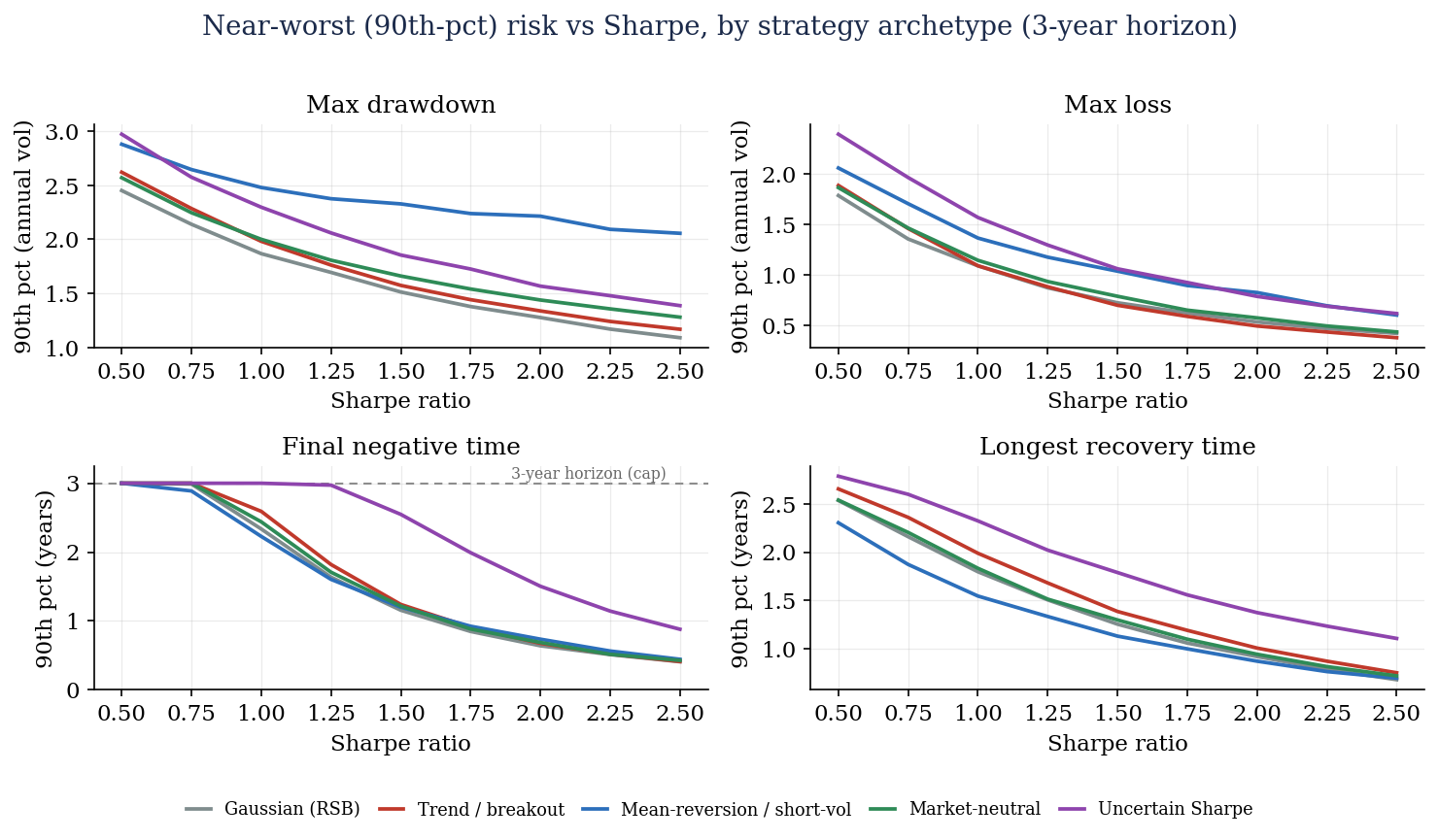}
\caption{Near-worst (90th-percentile) value of each risk measure as a function of the Sharpe ratio,
by archetype, over a three-year horizon. Differences are largest precisely where risk is largest, at
low Sharpe. \emph{Final negative time} is right-censored at the horizon: the flat segment at $3$ years
means that, at low Sharpe, at least one path in ten is still under water at the end of the window --- a
real outcome, not an artefact.}
\label{p2:fig:grid}
\end{figure}

Figure~\ref{p2:fig:grid} extends the comparison across Sharpe ratios. The archetype ordering is stable:
the gaps are widest at low Sharpe, where drawdown risk is largest in absolute terms and the
penalty for the wrong distributional assumption is greatest.

\subsubsection{Uncertainty about the Sharpe exceeds style}
A second effect not only rivals but exceeds the distributional ones, and comes not from the shape of
returns but from our ignorance of the drift. Marginalising over a realistic Sharpe standard error lifts
the near-worst maximum loss to $1.48\times$ the Gaussian benchmark, recovery time to $1.30\times$, final
negative time to $1.27\times$ and maximum drawdown to $1.23\times$ --- larger than any single style
effect in this study, and broader, since it pushes every measure up at once. The practical corollary reinforces the
original RSB message from the opposite direction: much of what looks like an ``abnormal'' drawdown is
simply the admission that a Sharpe estimated on a few years of data was never as precise as its single
decimal suggested.

\FloatBarrier

\section{Long memory: fractional Brownian motion}\label{sec:longmemory}
\subsection{From short memory to long memory}
The engine of Section~4 added persistence through an AR(1) term, whose autocorrelation
decays geometrically: memory that fades within a handful of days. Real strategy P\&L can instead show
\emph{long} memory --- autocorrelation that decays as a power law. The canonical model is fractional
Brownian motion~\cite{mvn1968}, a Gaussian process $B_H(t)$ whose increments (fractional Gaussian noise)
carry a single parameter, the Hurst exponent $H\in(0,1)$:
\begin{equation}
\mathbb{E}[B_H(t)^2]= t^{2H},\qquad
\rho(k)=\tfrac12\big(|k{-}1|^{2H}-2|k|^{2H}+|k{+}1|^{2H}\big).
\end{equation}
$H=\tfrac12$ is ordinary Brownian motion with independent increments; $H>\tfrac12$ is \emph{persistent}
(same-direction moves cluster; long runs and long droughts); $H<\tfrac12$ is \emph{anti-persistent}
(mean-reverting increments). Jean-Philippe Bouchaud suggested that the persistent case is relevant for
statistical-arbitrage books --- the motivation for this note.

\subsection{Model and normalisation}
We simulate fractional Gaussian noise exactly by the Davies--Harte circulant-embedding method (vectorised
across paths), standardise it to pooled unit variance, and impose the drift by the same affine transform
used throughout the series,
\begin{equation}
r_t \;=\; \frac{\Sigma}{\sqrt{250}}\left(\frac{S}{\sqrt{250}}+\tilde y_t^{(H)}\right),\qquad \Sigma=1 .
\label{p3:eq:tr}
\end{equation}
Equation~\eqref{p3:eq:tr} gives a daily mean $\Sigma S/250$ and a daily volatility $\Sigma/\sqrt{250}$, so the
\emph{one-day} Sharpe is $S/\sqrt{250}$. Thus $S=1$ is \emph{not} a daily Sharpe: it is the
Brownian-convention (annualised-style) Sharpe implied by the one-day mean and one-day volatility. What we
hold fixed across $H$ is the one-day mean and one-day volatility. Under fBm the cumulative standard
deviation at horizon $T$ (days) is $(\Sigma/\sqrt{250})\,T^{H}$, so the \emph{effective} $T$-day Sharpe,
\begin{equation}
\mathrm{SR}_T(H)=\frac{S\,T^{\,1-H}}{\sqrt{250}},
\end{equation}
\emph{declines} with $H$ for $H>\tfrac12$: persistence buys multi-horizon dispersion, not multi-horizon
edge. This is the crux of the whole note.

\subsubsection{One-day-fixed versus horizon-fixed risk}
Two normalisations answer two different questions. \textbf{One-day-fixed} (the convention above) is what a
manager actually quotes: identical daily risk, but --- because dispersion grows as $T^{H}$ ---
super-diffusive risk at multi-year horizons. \textbf{Horizon-fixed} instead rescales the stochastic increments by
$T^{1/2-H}$ so that the cumulative standard deviation at the three-year horizon matches the Brownian
($H=\tfrac12$) value; it removes the dispersion-scaling effect and isolates the residual influence beyond that scaling. Concretely, we rescale only the stochastic fGn innovation, leaving the daily drift unchanged, so terminal volatility \emph{and} terminal Sharpe are equalised across $H$. We report both conventions, and keep them clearly separate.

\subsection{Results}
\subsubsection{One-day-fixed: the multipliers are dominated by dispersion scaling}
\begin{figure}[htbp]\centering
\includegraphics[width=\linewidth]{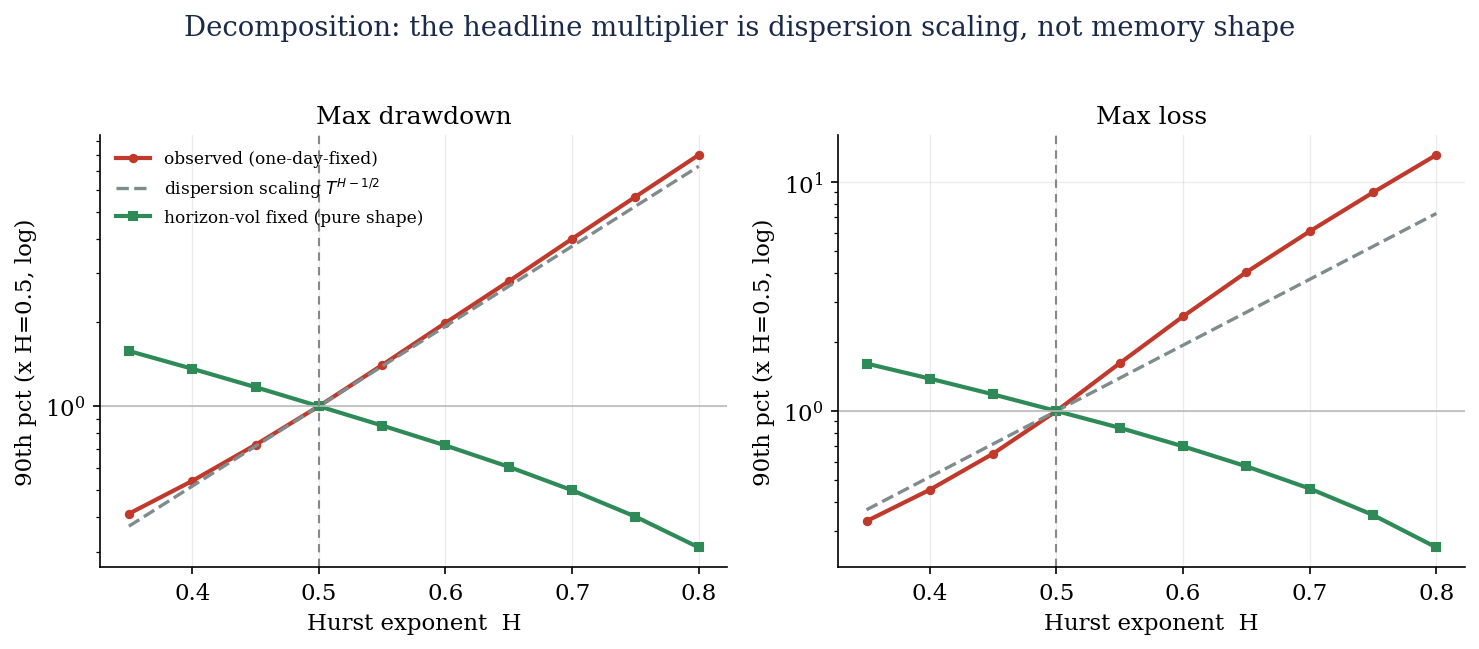}
\caption{Near-worst depth vs $H$ (log scale). The observed one-day-fixed multiplier (red) tracks the
dispersion-scaling line $T^{H-1/2}$ (grey dashed) almost exactly for maximum drawdown; the horizon-fixed
curve (green) --- which removes the scaling --- lies at or below one for persistent $H$, showing that the residual beyond scaling does not deepen drawdowns.}
\label{p3:fig:decomp}
\end{figure}

\begin{figure}[htbp]\centering
\includegraphics[width=\linewidth]{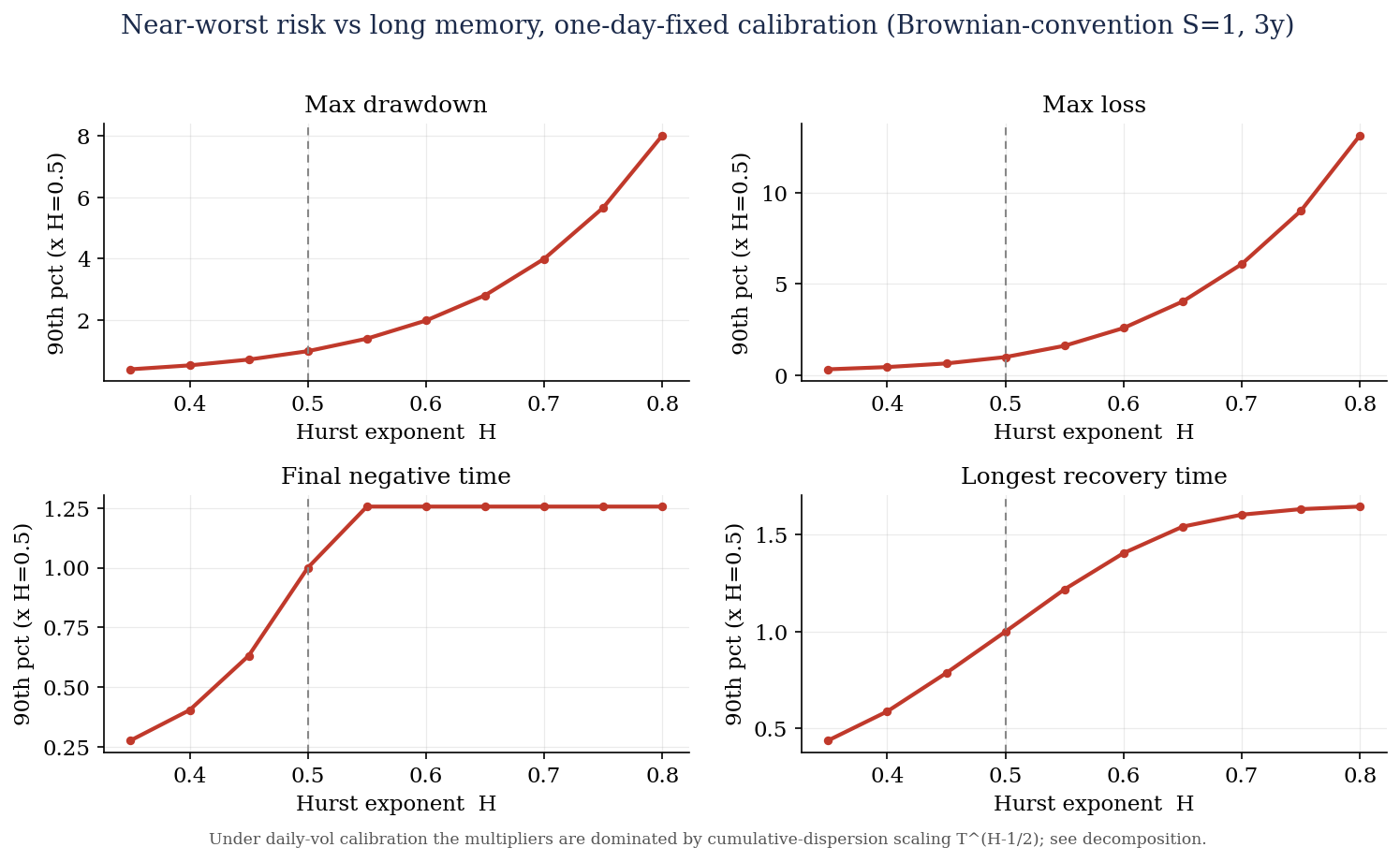}
\caption{One-day-fixed near-worst risk vs the Hurst exponent (Brownian-convention $S=1$, three years,
common random numbers across $H$). The steep rise is the cumulative-dispersion effect of \S3.1, not a residual
effect beyond dispersion scaling.}
\label{p3:fig:measures}
\end{figure}

Table~\ref{p3:tab:one} and Figure~\ref{p3:fig:measures} give the four decision measures at Brownian-convention
Sharpe $S=1$ over three years, as multiples of the Brownian ($H=\tfrac12$) benchmark. The depth measures
appear to explode --- near-worst maximum drawdown reaches $8\times$ and maximum loss $13\times$ at $H=0.8$.
But Table~\ref{p3:tab:decomp} and Figure~\ref{p3:fig:decomp} decompose each depth multiplier into the trivial
cumulative-dispersion factor $T^{H-1/2}$ and a residual component beyond dispersion scaling. For maximum drawdown the
residual is only $1.03$--$1.10$: essentially the \emph{entire} multiplier is dispersion scaling. Maximum
loss keeps a larger residual (up to $1.8\times$) --- the single worst excursion of a persistent path is
somewhat worse than scaling alone --- but even there the headline is mostly $T^{H-1/2}$. The honest reading
is that the $8$--$13\times$ figures measure the failure of $\sqrt{T}$ annualisation under persistence, not
an exotic deepening of path geometry.

\begin{table}[htbp]\centering
\caption{One-day-fixed near-worst (90th-percentile) risk by Hurst exponent, as a multiple of the Brownian
($H=0.5$) benchmark (Brownian-convention $S=1$, three years). ``$\,^\dagger$'' marks a statistic
right-censored at the 3-year horizon.}
\label{p3:tab:one}
\small\renewcommand{\arraystretch}{1.2}\setlength{\tabcolsep}{9pt}
\begin{tabular}{lcccc}
\toprule
\rowcolor{ink!12}\textbf{Hurst }$H$ & \textbf{max drawdown} & \textbf{max loss} & \textbf{final neg.\ time} & \textbf{longest recovery} \\
\midrule
0.40 (anti-persist.) & $0.54\times$ & $0.45\times$ & $0.40\times$ & $0.59\times$ \\
0.50 (Brownian)      & $1.00\times$ & $1.00\times$ & $1.00\times$ & $1.00\times$ \\
0.60                 & $1.99\times$ & $2.59\times$ & $1.26\times$ & $1.41\times$ \\
0.70                 & $4.00\times$ & $6.09\times$ & $1.26\times^{\dagger}$ & $1.60\times$ \\
0.80 (persistent)    & $8.01\times$ & $13.13\times$ & $1.26\times^{\dagger}$ & $1.65\times$ \\
\bottomrule
\end{tabular}
\end{table}

\begin{table}[htbp]\centering
\caption{Decomposition of the depth multipliers into cumulative-dispersion scaling $T^{H-1/2}$ and a
residual effect beyond dispersion scaling ($=\text{observed}/T^{H-1/2}$), $T=750$ days. For maximum drawdown the residual
is near unity: the multiplier is almost pure scaling.}
\label{p3:tab:decomp}
\small\renewcommand{\arraystretch}{1.2}\setlength{\tabcolsep}{10pt}
\begin{tabular}{lccccc}
\toprule
\rowcolor{ink!12}\textbf{Hurst }$H$ & $T^{H-1/2}$ & \textbf{max DD obs.} & \textbf{max DD residual} & \textbf{max loss obs.} & \textbf{max loss residual} \\
\midrule
0.40 & $0.52$ & $0.54\times$ & $1.04$ & $0.45\times$ & $0.88$ \\
0.60 & $1.94$ & $1.99\times$ & $1.03$ & $2.59\times$ & $1.34$ \\
0.70 & $3.76$ & $4.00\times$ & $1.06$ & $6.09\times$ & $1.62$ \\
0.80 & $7.29$ & $8.01\times$ & $1.10$ & $13.13\times$ & $1.80$ \\
\bottomrule
\end{tabular}
\end{table}

\subsubsection{Horizon-fixed: persistence does not deepen drawdowns}
Table~\ref{p3:tab:hor} repeats the experiment with the cumulative dispersion pinned at the three-year horizon
(so all $H$ share the same terminal volatility and the same terminal Sharpe). The picture inverts: for
persistent $H$ the depth measures fall \emph{below} the Brownian benchmark ($0.72$, $0.50$, $0.31\times$ for
maximum drawdown at $H=0.6,0.7,0.8$), because a persistent path of a given terminal spread trends more
smoothly and reverses less; it is the \emph{anti-persistent} paths that, zig-zagging to the same endpoint,
produce deeper interior drawdowns ($1.36\times$). Equalising the terminal dispersion does not equalise it at intermediate dates: the interior cumulative variance scales as $(t/T)^{2H-1}$ relative to Brownian --- below it for persistent $H$, above it for anti-persistent $H$ --- mechanically shrinking interior drawdowns for $H>\tfrac12$ and enlarging them for $H<\tfrac12$. There is thus no ``pure memory'' deepening of depth to
report --- the entire depth story of \S3.1 is the multi-horizon dispersion effect.

\begin{table}[htbp]\centering
\caption{Horizon-fixed near-worst (90th-percentile) risk: cumulative volatility equalised across $H$ at the
three-year horizon (multiples of the $H=0.5$ value). Isolating the residual beyond dispersion scaling, persistence \emph{reduces}
depth; anti-persistence increases it.}
\label{p3:tab:hor}
\small\renewcommand{\arraystretch}{1.2}\setlength{\tabcolsep}{9pt}
\begin{tabular}{lcccc}
\toprule
\rowcolor{ink!12}\textbf{Hurst }$H$ & \textbf{max drawdown} & \textbf{max loss} & \textbf{final neg.\ time} & \textbf{longest recovery} \\
\midrule
0.40 & $1.36\times$ & $1.38\times$ & $1.10\times$ & $1.03\times$ \\
0.50 & $1.00\times$ & $1.00\times$ & $1.00\times$ & $1.00\times$ \\
0.60 & $0.72\times$ & $0.70\times$ & $0.88\times$ & $0.96\times$ \\
0.70 & $0.50\times$ & $0.46\times$ & $0.77\times$ & $0.89\times$ \\
0.80 & $0.31\times$ & $0.25\times$ & $0.60\times$ & $0.81\times$ \\
\bottomrule
\end{tabular}
\end{table}

\paragraph{Censoring of the time measures.} Final negative time and longest recovery are right-censored at
the three-year horizon; the reported 90th percentile is a lower bound. Under one-day-fixed calibration the
fraction of paths still below zero at three years rises from $4\%$ at $H=0.5$ to $33\%$ at $H=0.7$ and
$40\%$ at $H=0.8$ --- again chiefly the consequence of the declining effective horizon Sharpe
$\mathrm{SR}_T(H)$, not of path geometry.

\FloatBarrier
\subsection{Validation}
For driftless Brownian motion the fraction of time the cumulative sum stays positive follows L\'evy's
arcsine law, $\Pr[F\le x]=\tfrac{2}{\pi}\arcsin\sqrt{x}$; for fBm this generalises, with the same mean
($\tfrac12$) but a variance that grows with $H$~\cite{sadhu2017}. Our simulated positive-time fraction matches
this: at $H=0.5$ the quantiles $\{0.024,0.147,0.503,0.856,0.976\}$ equal the arcsine values
$\{0.024,0.146,0.500,0.854,0.976\}$ to three decimals, and the variance rises $0.107\!\to\!0.126\!\to\!0.149$
at $H=0.3,0.5,0.7$. Two further checks: an independent exact-Cholesky generator reproduces the depth
multipliers (e.g.\ $1.98$ vs $2.03$ at $H=0.6$; $4.00$ vs $4.08$ at $H=0.7$), and the Kolmogorov--Smirnov
distance to the continuous arcsine law shrinks with step count ($0.037\!\to\!0.010$ from $n=250$ to
$4000$), a finite-step correction converging to the continuum. These validate the driftless
\emph{occupation-time} statistics of the simulated fBm --- a sanity check on the engine --- not the
drawdown-with-drift distributions themselves, for which no closed form is claimed.

\section{Synthesis: every process assumption on one axis}\label{sec:synthesis}

\begin{figure}[htbp]\centering
\includegraphics[width=\linewidth]{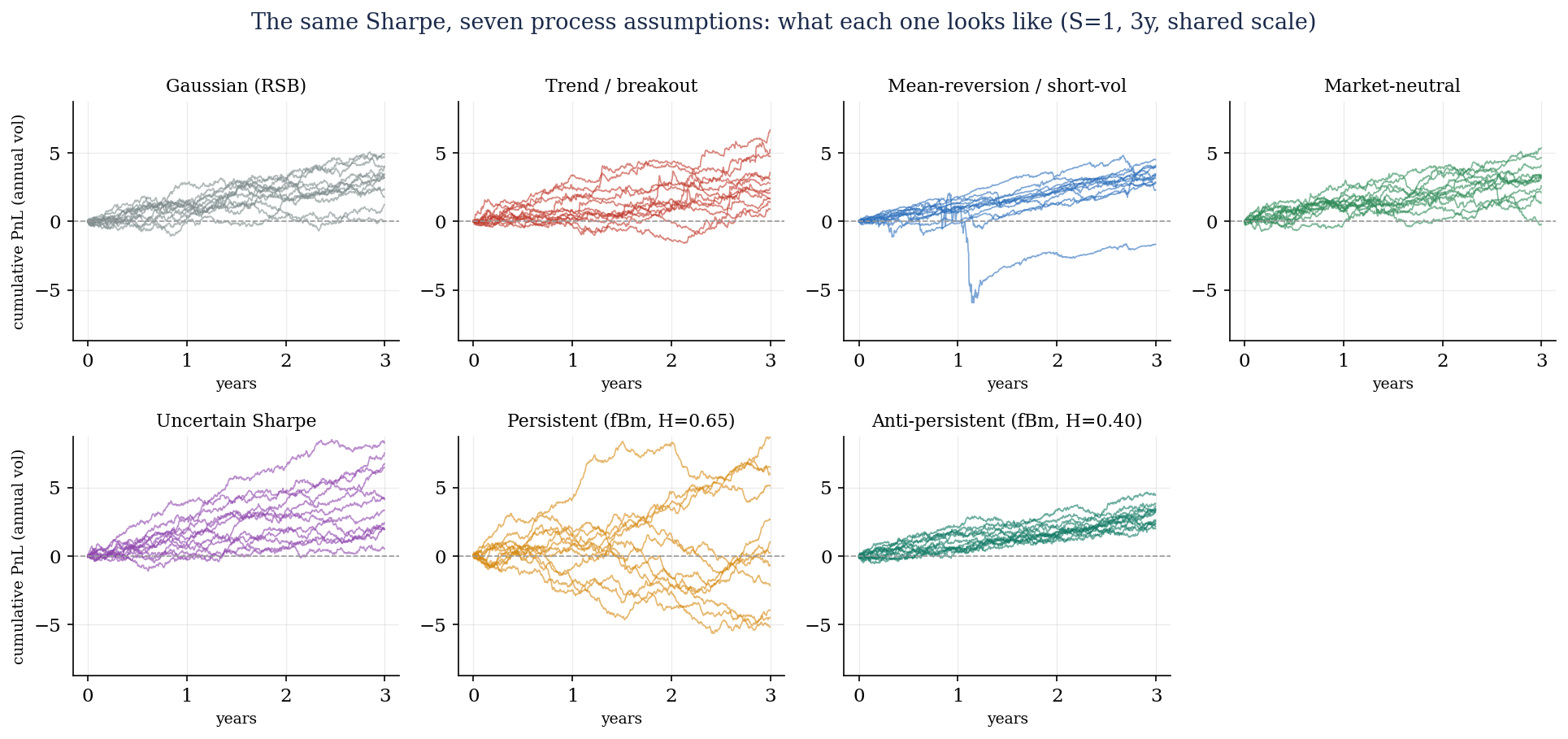}
\caption{Twelve three-year equity lines for each of the seven process assumptions, all with the same
Brownian-convention Sharpe ($S=1$) and daily volatility, on a shared vertical scale. The persistent (fBm)
panel visibly disperses the most --- the manifestation of the scaling effect quantified in
Table~\ref{p3:tab:master}.}
\label{p3:fig:gallery}
\end{figure}

It is worth placing long memory beside the other departures from the Brownian benchmark examined in this
series. Table~\ref{p3:tab:master} collects the near-worst (90th-percentile) risk of all seven process
assumptions --- the four archetypes of Section~4, the Sharpe-uncertainty layer, and persistent and
anti-persistent fractional Brownian motion --- on a single, consistent \emph{one-day-fixed} basis (identical
daily mean and volatility, Brownian-convention $S=1$), each as a multiple of the Gaussian (RSB) benchmark.
Figure~\ref{p3:fig:gallery} shows what each one looks like.

Two readings stand out. First, under the one-day-fixed convention the largest multiplier in the table comes from persistent fBm --- because it changes the daily-to-horizon scaling itself: at
$H=0.65$ the near-worst maximum drawdown and maximum loss reach $2.8\times$ and $4.0\times$ the Gaussian,
dwarfing the $5$--$50\%$ distortions of skew, fat tails, volatility clustering and Sharpe uncertainty.
Second --- the honest caveat --- for the fBm rows that magnitude is predominantly the cumulative-dispersion
scaling of \S3.1 (the failure of $\sqrt{T}$ annualisation), whereas the archetype effects are genuine
\emph{shape} effects at essentially unchanged dispersion. The two mechanisms differ; the table makes them
comparable only because they share the daily-volatility convention a manager actually quotes.

\begin{table}[htbp]\centering
\caption{\textbf{Every process assumption on one axis.} Near-worst (90th-percentile) risk of the four Part-2 archetypes, the Sharpe-uncertainty layer, and persistent / anti-persistent fBm, all on a consistent \emph{one-day-fixed} basis (identical daily mean and volatility, Brownian-convention $S=1$, three years), as a multiple of the Gaussian (RSB) benchmark. Green $=$ milder than Gaussian, red $=$ worse. For the fBm rows the multiplier is predominantly the cumulative-dispersion scaling of \S3.1, not a shape effect.}
\label{p3:tab:master}
\renewcommand{\arraystretch}{1.35}\setlength{\tabcolsep}{9pt}\small
\begin{tabular}{l|cccc}
\arrayrulecolor{gray!50}\hline
\rowcolor{ink}\textcolor{white}{\textbf{~Process assumption}} & \textcolor{white}{\textbf{Max drawdown}} & \textcolor{white}{\textbf{Max loss}} & \textcolor{white}{\textbf{Final negative time}} & \textcolor{white}{\textbf{Longest recovery time}}\\
\hline
~Gaussian (RSB) & \cellcolor{gray!12}1.00$\times$ & \cellcolor{gray!12}1.00$\times$ & \cellcolor{gray!12}1.00$\times$ & \cellcolor{gray!12}1.00$\times$\\
~Trend / breakout & \cellcolor[HTML]{F4EEEF}1.06$\times$ & \cellcolor[HTML]{F4F4F5}1.01$\times$ & \cellcolor[HTML]{F4E9E9}1.10$\times$ & \cellcolor[HTML]{F4E8E8}1.11$\times$\\
~Mean-reversion / short-vol & \cellcolor[HTML]{F4CCC9}1.34$\times$ & \cellcolor[HTML]{F4D0CE}1.30$\times$ & \cellcolor[HTML]{EEF4F1}0.96$\times$ & \cellcolor[HTML]{E2F1E4}0.88$\times$\\
~Market-neutral & \cellcolor[HTML]{F4EBEC}1.08$\times$ & \cellcolor[HTML]{F4EDEE}1.07$\times$ & \cellcolor[HTML]{F4F1F3}1.03$\times$ & \cellcolor[HTML]{F4F2F4}1.02$\times$\\
~Uncertain Sharpe & \cellcolor[HTML]{F4DBDA}1.21$\times$ & \cellcolor[HTML]{F4C0BC}1.43$\times$ & \cellcolor[HTML]{F4D3D1}1.28$\times$ & \cellcolor[HTML]{F4D1CF}1.29$\times$\\
~Persistent (fBm, $H=0.65$) & \cellcolor[HTML]{F4B2AC}2.83$\times$ & \cellcolor[HTML]{F4B2AC}4.00$\times$ & \cellcolor[HTML]{F4D3D1}1.28$\times$ & \cellcolor[HTML]{F4B2AC}1.55$\times$\\
~Anti-persistent (fBm, $H=0.40$) & \cellcolor[HTML]{D6EED6}0.55$\times$ & \cellcolor[HTML]{D6EED6}0.45$\times$ & \cellcolor[HTML]{D6EED6}0.41$\times$ & \cellcolor[HTML]{D6EED6}0.60$\times$\\
\hline
\end{tabular}
\end{table}

The gallery makes the intuition physical. Same headline Sharpe, seven different lived experiences of risk:
the Gaussian drifts noisily; trend grinds then jumps; the short-vol book climbs smoothly then cliffs;
market-neutral hugs the benchmark; Sharpe uncertainty fans the outcomes wide; the persistent book wanders
far in one direction for years at a time; the anti-persistent book barely leaves its trend.

\FloatBarrier

\section{Practical calibration and discussion}\label{sec:practical}
The preceding results are only useful if a manager can act on them. This section turns them into a calibration recipe and is explicit about scope.

\subsection{Practical implications}
Table~\ref{p2:tab:archlookup} gives revised, archetype-specific near-worst thresholds. A risk manager
should select the archetype matching the manager's style and read the relevant cell: a live reading
beyond it is the evidence-based trigger to revise the Sharpe downward or to cut --- replacing the
single Gaussian table with one calibrated to how that style actually behaves.

\begin{table}[ht]\centering
\caption{\textbf{The archetype decision panel.} Near-worst (90th-percentile) risk by strategy archetype and assumed Sharpe ratio, over a three-year horizon. For a manager whose style matches an archetype, these supersede the single Gaussian table: a live reading beyond the relevant cell is the signal to revise the Sharpe or cut. Cells shade from calm green to alarm red by severity within each measure. Depth in annual-vol units; time in years.}
\label{p2:tab:archlookup}
\renewcommand{\arraystretch}{1.5}\setlength{\tabcolsep}{10pt}\small
\begin{tabular}{ll|cccc}
\arrayrulecolor{gray!50}\hline
\rowcolor{ink} & \textcolor{white}{\textbf{Archetype}} & \textcolor{white}{\textbf{S=0.5}} & \textcolor{white}{\textbf{S=1.0}} & \textcolor{white}{\textbf{S=1.5}} & \textcolor{white}{\textbf{S=2.0}}\\
\hline
\multicolumn{6}{l}{\rule{0pt}{2.6ex}\textbf{\textcolor{ink}{Max drawdown}}~\textnormal{\footnotesize(annual vol)}}\\[1pt]
 & Gaussian (RSB) & \cellcolor[HTML]{FAD4B9}2.45 & \cellcolor[HTML]{F4F0C9}1.87 & \cellcolor[HTML]{E2EFD1}1.51 & \cellcolor[HTML]{D6EED6}1.28\\
 & Trend / breakout & \cellcolor[HTML]{F8C6B4}2.62 & \cellcolor[HTML]{FAF1C6}1.98 & \cellcolor[HTML]{E5EFCF}1.57 & \cellcolor[HTML]{D9EED5}1.34\\
 & Mean-reversion / short-vol & \cellcolor[HTML]{F4B2AC}2.88 & \cellcolor[HTML]{F9D1B8}2.48 & \cellcolor[HTML]{FCDDBD}2.33 & \cellcolor[HTML]{FDE6C0}2.21\\
 & Market-neutral & \cellcolor[HTML]{F8CAB5}2.57 & \cellcolor[HTML]{FBF1C6}2.00 & \cellcolor[HTML]{EAEFCD}1.66 & \cellcolor[HTML]{DEEFD2}1.44\\
\addlinespace[2pt]\hline
\multicolumn{6}{l}{\rule{0pt}{2.6ex}\textbf{\textcolor{ink}{Max loss}}~\textnormal{\footnotesize(annual vol)}}\\[1pt]
 & Gaussian (RSB) & \cellcolor[HTML]{F8C8B4}1.79 & \cellcolor[HTML]{F5F0C8}1.09 & \cellcolor[HTML]{E2EFD1}0.73 & \cellcolor[HTML]{D8EED5}0.54\\
 & Trend / breakout & \cellcolor[HTML]{F6C0B1}1.88 & \cellcolor[HTML]{F5F0C8}1.09 & \cellcolor[HTML]{E1EFD1}0.70 & \cellcolor[HTML]{D6EED6}0.50\\
 & Mean-reversion / short-vol & \cellcolor[HTML]{F4B2AC}2.06 & \cellcolor[HTML]{FEEAC1}1.37 & \cellcolor[HTML]{F2F0CA}1.04 & \cellcolor[HTML]{E7EFCE}0.83\\
 & Market-neutral & \cellcolor[HTML]{F7C2B2}1.86 & \cellcolor[HTML]{F8F0C7}1.15 & \cellcolor[HTML]{E5EFCF}0.79 & \cellcolor[HTML]{DAEED4}0.58\\
\addlinespace[2pt]\hline
\multicolumn{6}{l}{\rule{0pt}{2.6ex}\textbf{\textcolor{ink}{Final negative time}}~\textnormal{\footnotesize(years)}}\\[1pt]
 & Gaussian (RSB) & \cellcolor[HTML]{F4B2AC}3.00 & \cellcolor[HTML]{FAD5BA}2.34 & \cellcolor[HTML]{E8EFCE}1.16 & \cellcolor[HTML]{D6EED6}0.64\\
 & Trend / breakout & \cellcolor[HTML]{F4B2AC}3.00 & \cellcolor[HTML]{F8C8B4}2.59 & \cellcolor[HTML]{EBF0CD}1.24 & \cellcolor[HTML]{D7EED5}0.68\\
 & Mean-reversion / short-vol & \cellcolor[HTML]{F4B2AC}3.00 & \cellcolor[HTML]{FBDBBC}2.23 & \cellcolor[HTML]{E9EFCD}1.20 & \cellcolor[HTML]{D9EED5}0.74\\
 & Market-neutral & \cellcolor[HTML]{F4B2AC}3.00 & \cellcolor[HTML]{F9D0B7}2.44 & \cellcolor[HTML]{EAEFCD}1.22 & \cellcolor[HTML]{D8EED5}0.69\\
\addlinespace[2pt]\hline
\multicolumn{6}{l}{\rule{0pt}{2.6ex}\textbf{\textcolor{ink}{Longest recovery time}}~\textnormal{\footnotesize(years)}}\\[1pt]
 & Gaussian (RSB) & \cellcolor[HTML]{F5BAAF}2.54 & \cellcolor[HTML]{FFEFC3}1.80 & \cellcolor[HTML]{E8EFCE}1.25 & \cellcolor[HTML]{D8EED5}0.92\\
 & Trend / breakout & \cellcolor[HTML]{F4B2AC}2.66 & \cellcolor[HTML]{FCE1BE}1.99 & \cellcolor[HTML]{EEF0CC}1.38 & \cellcolor[HTML]{DCEED3}1.00\\
 & Mean-reversion / short-vol & \cellcolor[HTML]{F8CBB5}2.30 & \cellcolor[HTML]{F5F0C8}1.54 & \cellcolor[HTML]{E2EFD1}1.13 & \cellcolor[HTML]{D6EED6}0.87\\
 & Market-neutral & \cellcolor[HTML]{F5BAAF}2.54 & \cellcolor[HTML]{FEECC2}1.83 & \cellcolor[HTML]{EAEFCD}1.30 & \cellcolor[HTML]{D9EED5}0.94\\
\hline
\end{tabular}

\vspace{3pt}
{\footnotesize\centering \colorbox[HTML]{D6EED6}{\strut~calm~}\,$\rightarrow$\,\colorbox[HTML]{FFF1C4}{\strut~watch~}\,$\rightarrow$\,\colorbox[HTML]{F4B2AC}{\strut~alarm~}\hspace{1.5em}shade = severity within each measure\par}
\end{table}

The honest limitation is that realism costs universality: once skew, tails, clustering and persistence
enter, there is no single table that fits all strategies. The remedy is to calibrate to the manager's
\emph{own} returns. Two routes are available. The \textbf{parametric} route fits the archetype
parameters --- skew, kurtosis, GARCH coefficients and autocorrelation --- to the historical daily P\&L
and regenerates the tables. The \textbf{non-parametric} route, which a risk manager will trust more
because it assumes almost nothing, is a \emph{stationary block bootstrap}: resample blocks of the
actual return series (with a block length long enough to preserve autocorrelation and volatility
clustering), accumulate synthetic equity lines, and read the drawdown quantiles directly. The
parametric route can extrapolate into tails not yet observed but carries model risk; the bootstrap
cannot extrapolate but is assumption-light. Reporting both, and the gap between them, is itself a
useful measure of how fragile the risk estimate is.

\subsection{Relevance to statistical arbitrage --- conditional, not automatic}
Statistical-arbitrage returns are a plausible home for positive long memory: order flow is strongly
persistent~\cite{bgpw2004,lillofarmer2004}, crowded factors deleverage slowly, and signals decay over
weeks. But long memory in order flow or order signs does \emph{not} by itself imply long memory in realised
net P\&L. The relevance of this note to a given book is therefore conditional and empirical: \emph{when} a
strategy's P\&L increments pass persistence diagnostics, a drawdown table calibrated to daily volatility ---
implicitly assuming $H=\tfrac12$ --- can understate multi-year depth by the $T^{H-1/2}$ factor. The table
should not be applied merely because a strategy is labelled ``statistical arbitrage''.

\subsection{Practical use: $H$ as a stress axis}
Estimating $H$ from roughly $750$ daily observations is noisy and biased; short-memory autocorrelation,
volatility clustering, return smoothing and regime breaks can all masquerade as long memory, and standard
rescaled-range analysis is known to confound them (hence Lo's modified R/S). We therefore recommend
treating $H$ as a \emph{stress-testing axis}, not a plug-in point estimate: report near-worst thresholds
across a scenario set such as $H=0.5,0.6,0.7$, with uncertainty bands, using several diagnostics rather than
a single R/S or DFA number. Persistence diagnostics, so caveated, belong alongside Sharpe and volatility on
the risk report --- because they decide whether daily-calibrated drawdown limits mean what they appear to.

\FloatBarrier

\section{Conclusion}
The Brownian benchmark of Rej, Seager and Bouchaud is the right place to start and the wrong place to stop. Reframed as a Monte-Carlo experiment and extended to four decision measures, it already turns the keep-or-kill call into an evidence-based one. Relaxing the Gaussian assumption shows that those measures diverge with the strategy's style: payoff asymmetry decides whether the binding constraint is depth or duration, and uncertainty about the true Sharpe, the single largest effect in the non-Gaussian study, lifts every worst-case measure at once. Pushing persistence to fractional Brownian motion sharpens the picture rather than complicating it: the dramatic multi-year drawdowns of a persistent book are, for maximum-drawdown depth, almost entirely the self-similar growth of cumulative dispersion as $T^{H}$ rather than $\sqrt{T}$. Long memory is dangerous not because it reshapes the geometry of drawdowns but because it breaks the link between daily risk and horizon risk; equalise the horizon dispersion and persistence is, if anything, milder.

The practical lesson is uniform across the analysis: the Sharpe ratio is necessary but badly insufficient for judging when a live strategy has crossed from normal pain into genuine trouble, and a risk limit built from daily volatility can miss the difference entirely. Risk tables should be calibrated to the strategy's style and, where the data allow, to its own history, reporting skew, tails, clustering and a persistence (Hurst) diagnostic alongside the Sharpe. The honest limitation is that realism costs universality: the archetypes are stylised, the fractional Brownian model is stationary and Gaussian by construction, and the long-memory results are documented at a single three-year horizon. Each is a natural direction for further work; none overturns the central message, that when the process departs from Brownian motion, daily-volatility calibration quietly understates the risk that matters.

\section*{Acknowledgements}
The long-memory extension of Section~5 was prompted by a suggestion from Jean-Philippe Bouchaud.
AI-assisted tools supported parts of the coding, validation, drafting and editorial process; the author
directed and independently checked the analysis and takes full responsibility for all modelling choices,
results, interpretations and conclusions.

\end{document}